\documentclass[11pt,a4paper,fleqn,usenatbib,useAMS,]{mnras}
\usepackage[T1]{fontenc}
\usepackage{ae,aecompl}
\usepackage{mathptmx}

\usepackage{graphicx}	
\usepackage{amsmath}	
\usepackage{amssymb}	
\usepackage{multicol}
\usepackage{pdflscape}	
\usepackage{placeins}
\usepackage{float}
\usepackage{xcolor}
\usepackage{enumitem}
\usepackage{ulem}
\usepackage{soul}
\usepackage{hyperref}
\usepackage{url}
\usepackage{breqn}
\usepackage{gensymb}

\title[Compact symmetric objects]{X-ray timing and spectral characteristics of compact symmetric objects}

\author[Subhashree Swain et al.]{
Subhashree Swain$^{1,2}$,
\thanks{E-mail: subhashree.swain@iucaa.in}
C. S. Stalin$^{2}$,
Vaidehi S. Paliya$^{1}$,
D. J. Saikia$^{1}$,
\\
$^{1}$ \textit{Inter-University Centre for Astronomy and Astrophysics (IUCAA), SPPU Campus, Pune 411~007, India} \\
$^{2}$ \textit{Indian Institute of Astrophysics, Block II, Koramangala, Bangalore 560~034, India }
}

\date{Accepted 2024 December 20. Received YYY; in original form ZZZ}

\begin{document}

\label{firstpage}
\pagerange{\pageref{firstpage}--\pageref{lastpage}}
\maketitle

\begin{abstract}
Compact Symmetric Objects (CSOs) are a distinct category of jetted active
galactic nuclei whose high-energy emission is not well understood. We examined the X-ray characteristics of 17 bona fide CSOs 
using observations from {\it Chandra}, {\it XMM-Newton} and {\it NuSTAR}. Among the sources with {\it XMM-Newton} observations, we found two sources, J0713+4349 and J1326+3154 to show clear evidence of variations in the soft (0.3$-$2 keV), 
the hard (2$-$10 keV) and the total energy (0.3$-$10 keV) bands with the 
normalised excess variance (F$_{var}$) as large as 1.17$\pm$0.27. 
Also, the F$_{var}$ is found to be larger in the hard band relative to the 
soft band for J1326+3154. From the analysis of the hardness ratio (HR) with count rate, we found both sources to show a harder when brighter (HWB) trend. Similarly, 
in the Chandra observations, we found one source, J0131+5545, to show flux variations in the total energy band (0.5$-$7 keV). We discuss possible reasons for about 82 per cent of the CSOs being non-variable. From spectral analysis, carried out in a 
homogeneous manner, we found the existence of obscured as well as unobscured 
CSOs. Three CSOs, J0111+3906, J1407+2827 and J2022+6136, were found to have the intrinsic neutral hydrogen column density N$_{\rm H,z} > 10^{23}$ cm$^{-2}$, consistent with earlier analyses. For the majority of the CSOs, the observed hard X-ray emission is expected to be dominated by their mildly relativistic jet emission. For the sources, J0713+4349, J1347+1217, J1407+2827, J1511+0518 and J2022+6136, the confirmed detection of Fe K$\alpha$ emission line suggests a significant contribution from the disk/corona. Our results point to diverse X-ray characteristics of CSOs. 
\end{abstract}




\section{Introduction}


Compact symmetric objects (CSOs) are active galactic
nuclei (AGN) that are characterized by double compact lobes/hot
spots on opposite sides of an active nucleus or core, with an overall projected linear size smaller than about 1 kpc. The CSOs tend to have an inverted radio spectrum with a spectral peak at GHz frequencies, thereby being a sub-class of the peaked-spectrum radio sources \citep{Wilkinson1994, readhead1978, readhead1980, readhead1996}. 
In the radio band, their morphology is similar to those of classical
double-lobed radio sources, except that their sizes are up to about a few
thousand times smaller. Their reasonably symmetrical radio morphology, unlike the core-jet sources, and weak radio cores suggest that these are usually viewed at large angles to the line of sight.

The CSO classification is a more physically motivated classification than compact steep-spectrum (CSS) and peaked-spectrum (PS) sources \citep{Tremblay2016}. CSS sources are defined to be < 20 kpc, while PS sources have a turnover in the radio spectrum from MHz to GHz frequencies and a steep ($\alpha$ > 0.5; $S \propto \nu^{-\alpha}$) radio spectrum above the turnover frequency. PS sources are often contaminated by relativistic beaming, while CSS sources include a diverse range of AGN types. The small sizes of CSS and PS sources, which include the CSOs, could be due to (i) a dense environment in the innermost few tens to hundreds 
of parsecs from the radio core or AGN, which inhibits their advancement to form large sources; (ii) their young ages so that they evolve into larger sources as they age; or (iii) their transient or recurrent nuclear jet activity. While a combination of these may be playing a role in different sources, the large number of compact sources compared with their larger counterparts suggests that many are unlikely to evolve into large sources \citep[for a review, see][]{2021A&ARv..29....3O}. Recently, \cite{Readhead_2024} have suggested that CSOs are typically less than about 5000 yr old and do not evolve into larger sources. The kinematic ages of these compact sources estimated from the velocity of advancement of the hotspots, which has a median value of $\sim$0.1c, range from tens to several thousand years (\citealt{2002NewAR..46..263C,2021A&ARv..29....3O}, and references therein). Their radiative ages are also similar, suggesting that these are young objects.   

As these young, compact, luminous radio sources traverse outwards, they interact vigorously with the interstellar medium (ISM) of their host galaxies, driving shocks, heating the ISM and accelerating particles. Understanding such feedback processes on these small scales is also important as these could affect the feeding of the AGN and star formation in the circumnuclear region, thereby affecting the evolution of their host galaxies. High energy X-ray emission could help understand the physical conditions in these sources, their interactions with the environment and constrain models of their evolution. The X-ray emission could arise from the corona above the accretion disk, the lobes and jets of the radio source and hot gas shocked by the advancing radio source. 
However, the number of CSOs with quality X-ray observations
are limited. Also, the small numbers of sources
that have been studied in X-rays are faint with 
observed 2$-$10 keV flux values of 10$^{-14}$ - 10$^{-13}$
erg cm$^{-2}$ s$^{-1}$ \citep{Siemiginowska2016}. The limited
studies available in the literature indicate that these
sources have complex X-ray spectra with their primary X-ray
emission getting modified by reflection and absorption processes.
Analysis of the hard X-ray observations of a CSO, namely OQ+208
by \cite{2019ApJ...884..166S} indicates the X-ray emission in the source being produced either by the corona or the relativistic jet.
Recently, \cite{Sobolewska_2023} studied the X-ray variability on the time 
scale of years in two CSOs, namely, J1511+0518 and J2022+6136, using 
{\it XMM-Newton} and {\it Nuclear Spectroscopic Telescope Array (NuSTAR)} 
observations. In addition to finding
them variable on a time scale of years, they also found them to have intrinsic X-ray absorbing column densities (N$_{H,z}$) in excess of 10$^{23}$ cm$^{-2}$. 
X-ray studies of a large sample of CSOs are, therefore, needed to understand
their X-ray emission processes as well as their evolution.

 CSOs have also emerged as a new population of $\gamma$-ray emitting
AGN, with a few of them detected in the GeV band by the {\it Fermi} Gamma Ray
Space Telescope \citep{an2016, 2016ApJ...821L..31M, 2020A&A...635A.185P,2020ApJ...899..141L,2021MNRAS.507.4564P, 2022ApJS..263...24A, Gan_2022} and one among them namely, 
PKS 1413+135 is also detected in the 
TeV energy band \citep{2022ATel15161....1B}.

 Variability provides an important diagnostic for understanding the physical conditions in the nuclear regions of AGN 
\citep[e.g.][]{1993ARA&A..31..717M,1997ARA&A..35..445U,2006ASPC..360..111G}. Among radio-loud AGN, the most striking examples of variability are found in blazars where the jets are inclined at small angles to the line of sight. Although the Giga-Hertz Peaked Spectrum (GPS) and CSS sources, which include the CSOs, are expected to be inclined at larger angles to the line of sight,
short-term variability in X-rays provides a means of probing in the vicinity of the central engine i.e. the regions associated with the disk/corona and the relativistic jets. In the case of CSOs, the X-ray emission is often associated with 
processes occurring in the accretion flow or compact radio lobes. Some CSOs are also known for their Fe line emission and putative torus \citep{Siemiginowska2016, sobo2019, Sobolewska_2023}. The X-ray variation in some AGN could also be due to the movement of an intrinsic absorber along the line of sight on the timescales of months-to-years if it is part of a distant torus \citep[e.g][]{2002ApJ...571..234R, 2014MNRAS.437.1776M, 10.1093stad337} or days-to-weeks if located closer, in the broad-line region \citep[BLR; e.g.,][]{Risaliti_2005, puccetti_2007, svoboda_2015}. Hence short-term variability has the potential to probe the 
central regions of AGN.

Our goal in this work is to investigate the X-ray timing and spectral properties of a sample of CSO sources, and to study spectral trends with photon index, X-ray luminosity and Eddington ratio. We selected our sources from the catalogue of \cite{Kiehlmann_2024}. This catalogue contains a sample of 79 bona fide CSOs. These CSO sources have a projected linear size of less than  $\sim$800 pc. In Section 2, we summarize the details
of the final sample of CSOs selected for this study and the details of
the data and their reduction procedure. 
{In Section 3, we provide our results in three subsections concentrating on the
flux variability; spectral variability; and spectral modeling of the archival CSO data
performed to assess the X-ray luminosity and Eddington ratio in a
consistent manner.} The results are discussed in Section 4, while the conclusions are summarised in Section 5. 

\section{Sample and data reduction}
Our initial sample consists of 79 bona fide CSOs from 
\cite{Kiehlmann_2024}. The authors have taken care not to include any sources that are unlikely to be CSOs. Hence, this sample is well suited to understand their
nature. In this work, we utilized this bona fide sample to explore their
X-ray behaviour. A total of 26 out of the 79 sources have archival X-ray data available
from {\it XMM-Newton}, {\it Chandra} and {\it NuSTAR} observations. Restricting to sources with a net count larger than 100, we arrived at a final sample 
of 17 sources. Of the 17 sources, 13 sources have {\it XMM-Newton} observations, 
 8 sources have {\it Chandra} observations, 6 sources have {\it NuSTAR} observations. Out of them, 2 sources have  
{\it Chandra}, {\it XMM-Newton} and {\it NuSTAR} observations. The details of the 17 sources are given in Table \ref{table-1}. 
\subsection{{\it XMM-Newton}}
The details of the {\it XMM-Newton} observations are given in Table \ref{table-1}. 
We downloaded the observation data files from the HEASARC
archives. For reduction of the data, we used the {\it XMM-Newton} Science 
Analysis Software (SAS) version {\it 12.0.1}, and used data in the 0.3$-$10 keV energy band from only the EPIC-PN.
We applied the standard screening of the events, excluding the periods of flaring 
particle background. The resulting clean exposure time for each
source is listed in Table \ref{table-1} along with the net count.  
We extracted the source region of 40$\arcsec$ radius and background region of 40$\arcsec$ radius from the 
source-free region of the same CCD. As many of the sources were X-ray faint, 
we took care of the choice of flaring particle background rejection thresholds 
which optimize the signal-to-noise ratio of the final scientific products. We 
extracted 10 keV $<$ E $<$ 12 keV full field-of-view light 
curve as a monitoring tool of the background intensity with the threshold of 
0.4 counts/sec as recommended in the online documentation. Then, using the  filtered event lists and the corresponding Good 
time 
interval (GTI), we generated the spectra of the sources and background regions
which were then used to get the final spectra of the sources. 
 We binned the spectra to have at least 25 counts in each 
spectral bin and adopted the $\chi^2$ statistic for the goodness-of-fit test. To get the light curves using SAS packages, we used
{\it epiclccor} task with a binning of 600 sec. 
 We also generated the light curves with 300 sec and 1200 sec bins; however, we found that the shorter bins resulted in larger flux uncertainties, and larger bins may average-out short-time scale features, and the overall improvement in the obtained results was negligible. Hence, we choose 600 sec as bin size, which is optimal for our work. We followed the same bin size in {\it Chandra} and {\it NuSTAR} observations.

\subsection{{\it Chandra}}
The details of the {\it Chandra} observations are given in Table \ref{table-1}.
We utilized archival data in the 0.5$-$7 keV energy range from the Advanced CCD 
Imaging Spectrometer spectroscopy array (ACIS-S; \citealt{2003SPIE.4851...28G}). 
We used the level 2 data. The observations were made in the VFAINT model with 1/8 CCD 
readout to avoid pileup if sources were too bright. We used the {\it Chandra}
Interactive Analysis of Observations (CIAO) software version 4.15 and CALDB version 4.10.1
to re-process the level 2 event files. For timing analysis, we inspected the 
data using ds9 and made a circular region centered on the source with a radius 
of 2.5". For background, we used a nearby source-free region of radius 2.5" on the same CCD as the source. We extracted the light curves in
the 0.5$-$7 keV energy range using CIAO task {\it dmextract} with a
binning of 600 sec. For spectral analysis, we extracted the spectra and 
calibration files such as ARF and RMF with the CIAO script {\sc specextract}. 

\subsection{\it NuSTAR}
The details of Nuclear Spectroscopic Telescope Array ({\it NuSTAR}) \citep{harrison2013} observations are given in Table \ref{table-1}. We found the archival data for six CSOs only. 
{\it NuSTAR} has two identical co-aligned telescopes, each consisting of an
independent set of X-ray mirrors and a focal-plane detector, referred to as focal plane modules A and B (FPMA and FPMB)
that operate in the energy range 3–79 keV. The data reduction
was performed with {\sc nustardas} v1.8.6, available in the {\it NuSTAR}
Data Analysis Software. The event data files were calibrated
with the {\sc nupipeline} task using the response files from the Calibration Database caldb v.20180409 and HEASOFT version 6.25.
With the {\sc nuproducts} script, we generated both light curve and spectra of the source and
background. For both focal plane modules (FPMA, FPMB), we used a circular extraction region of radius 40" centered on the position of the
source. The background selection was made by taking a region free of sources with a 40" radius in the same detector quadrant. We extracted the light curve with binning of 600 sec in 3$-$20 keV band since the emission above 20 keV was found to be dominated by the background \citep[see also][]{sobo2019,Sobolewska_2023,Bronzini2024}. 
\begin{table*}
\centering
\caption[]{Details of the sources and their observations used in this 
work. Columns show the (1) J2000 name, (2,3) J2000 right ascension and declination, (4) redshift, z,  (5) the projected linear size in kpc, LS, (6) telescopes, (7) observation (Obs.) ID, (8) date of observation, (9) net exposure time, (10) net counts, (11) optical classification from Swain et al. (in preparation) : `Q' stands for quasar, `G' stands for galaxy. 
}
\label{table-1}
\small
\begin{tabular}{llllllllcrrr} 
\hline
Name &  RA & Dec & z & LS & Telescopes & OBSID & Date of Obs. & Exp. time & Net &Optical \\
&  &  & & (kpc) & & & &  (ks) & counts &type \\
(1)& (2) & (3) & (4)& (5)&(6) &(7) &(8) &  (9) & (10) &(11) \\
\hline
J0029+3456  & 00:29:14.24 & +34:56:32.25 & 0.517 & 0.180 & {\it XMM-Newton} & 0205180101 & 2004-01-08 & 10  &   572 & G \\
J0111+3906  & 01:11:37.32 & +39:06:28.10 & 0.668 & 0.056 & {\it XMM-Newton} & 0202520101 & 2004-01-09 & 13  &   146 & G\\
J0131+5545  & 01:31:13.82 & +55:45:12.98 & 0.036 & 0.016 & {\it Chandra}    & 21408      & 2019-03-29 & 5.7 &   424&G\\
J0713+4349  & 07:13:38.16 & +43:49:17.21 & 0.518 & 0.217 & {\it Chandra}    & 12845      & 2011-01-18 & 38  &  1754& G \\
            &             &              &       &       & {\it XMM-Newton} & 0202520201 & 2004-03-22 & 12  &  1259 & G\\
J1148+5924  & 11:48:50.36 & +59:24:56.36 & 0.011 & 0.012 & {\it Chandra}    & 10389      & 2009-07-20 & 39  &   408 &G \\
            &             &              &       &       & {\it NuSTAR} & 60601019002 & 2020-08-04 & 77 & 279 & \\
J1220+2916  & 12:20:06.82 & +29:16:50.72 & 0.002 & 0.002 & {\it Chandra}    & 7081 & 2007-02-22 & 122  &  3327 & G\\
            &             &              &       &       & {\it XMM-Newton} & 0205010101 & 2004-05-23 & 21  & 36178 & \\
J1326+3154  & 13:26:16.51 & +31:54:09.52 & 0.368 & 0.345 & {\it XMM-Newton} & 0502510301 & 2007-12-05 & 20  &   898 & G \\
J1347+1217  & 13:47:33.36 & +12:17:24.24 & 0.121 & 0.215 & {\it Chandra}    & 836        & 2000-02-24 & 25  &  1412 &Q\\
J1407+2827  & 14:07:00.40 & +28:27:14.69 & 0.077 & 0.016 & {\it Chandra}    & 16070      & 2014-09-04 & 35  &   713 & G\\
            &             &              &       &       & {\it XMM-Newton} & 0140960101 & 2003-01-31 & 10  &   711 & \\
             &             &              &       &       & {\it NuSTAR} & 60201043002 &2016-06-18 &51 &507 & \\
J1443+4044  & 14:42:59.32 & +40:44:28.94 & 2.593    &       & {\it XMM-Newton} & 0822530101 & 2019-01-18 & 33  &   601 & \\
J1511+0518  & 15:11:41.27 & +05:18:09.26 & 0.084 & 0.017 & {\it XMM-Newton} & 0822350101 & 2018-08-15 & 9.6 &   395 & G\\
             &             &              &       &       & {\it NuSTAR} & 60401024002 & 2019-01-08  & 70  &   248 & \\
J1609+2641  & 16:09:13.32 & +26:41:29.04 & 0.473 & 0.362 & {\it Chandra}    & 12846      & 2010-12-04 & 38  &   222 &G\\
J1723$-$6500  & 17:23:41.03 & -65:00:36.61 & 0.014 & 0.002 & {\it Chandra}    & 12849      & 2010-11-09 & 4.7 &   205 &G\\
            &             &              &       &       & {\it XMM-Newton} & 0845110101 & 2020-03-27 & 29  &  4622 &  \\
            &             &              &       &       & 
            {\it NuSTAR} &  60601020002 &2020-08-27 & 68&453  &  \\
J1939$-$6342  & 19:39:25.02 & -63:42:45.62 & 0.183 & 0.196 & {\it XMM-Newton} & 0784610201 & 2017-04-01 & 19  &  1194&G \\
J1945+7055  & 19:45:53.52 & +70:55:48.73 & 0.101 & 0.075 & {\it XMM-Newton} & 0784610101 & 2016-10-21 & 17  &   918 &G\\
J2022+6136  & 20:22:06.68 & +61:36:58.80 & 0.227 & 0.104 & {\it XMM-Newton} & 0784610301 & 2016-05-25 & 22  &  1029& G \\
&             &              &       &       & 
{\it NuSTAR} & 60401023002 & 2018-07-07 & 64  &  364 &  \\
J2327+0846  & 23:27:56.70 & +08:46:44.30 & 0.029 & 0.744 & {\it XMM-Newton} & 0200660101 & 2004-06-02 & 7.3 &   122 &G\\
&             &              &       &       & 
{\it NuSTAR} & 60001151002 & 2014-09-30 & 52  &  1388 &  \\
\hline
\end{tabular}
\end{table*}
\section{Results}
We provide the results in three subsections concentrating on the
flux variability, spectral variability and spectral modeling of the archival CSO data
performed to assess the X-ray luminosity and Eddington ratio in a
consistent manner.
\subsection{Flux variability}
To quantify the strength of X-ray flux variations in our sample of CSOs, we used
the quantity F$_{var}$, which is the square root of the normalised excess variance ($\sigma^2_{NXV}$) \citep{2002ApJ...568..610E,2003MNRAS.345.1271V} and is defined as 

\begin{equation}
F_{var} = \sqrt{\sigma^2_{NXV}} = \sqrt{\frac{S^2 - \bar{\sigma}^2_{err}}{\bar{x}^2}}
\end{equation}
where S is the sample variance of the light curve, $\bar{x}$ is the mean of 
x$_i$ measurements, and $\bar{\sigma}^2_{err}$ is mean square
error of each individual error $\sigma_{err,i}$ given by 
\begin{equation}
S^2 = \dfrac{1}{n-1} \sum_{i=1}^{n} (x_i - \bar{x})^2
\end{equation}

\begin{equation}
\bar{\sigma}^2_{err} = \dfrac{1}{n} \sum_{i=1}^{n} (\sigma_{err,i}^2)
\end{equation}

\begin{figure*}
\hbox{
\includegraphics[scale=0.55]{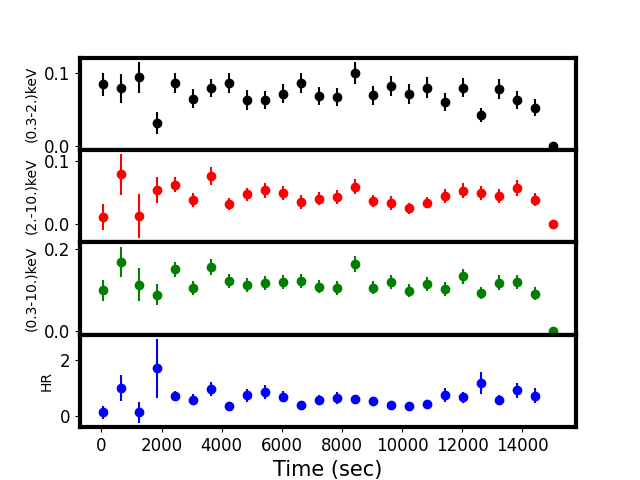}
\hspace*{-0.85cm}\includegraphics[scale=0.55]{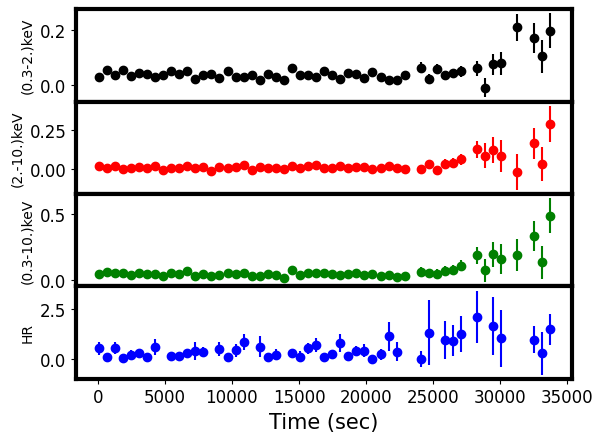}
     }
\caption{Light curves with a binning of 600 sec for the objects J0713+4349 
(left panel) and J1326+3154 (right panel) from observations with
{\it XMM-Newton}. In both the panels, from the top are shown the 
flux variations in the energy ranges of 0.3$-$2 keV, 2$-$10 keV and 
0.3$-$10 keV. The bottom panel shows the variation of HR.}
    \label{figure-1}
\end{figure*}

\begin{figure*}
\hbox{
	\includegraphics[scale=0.55]{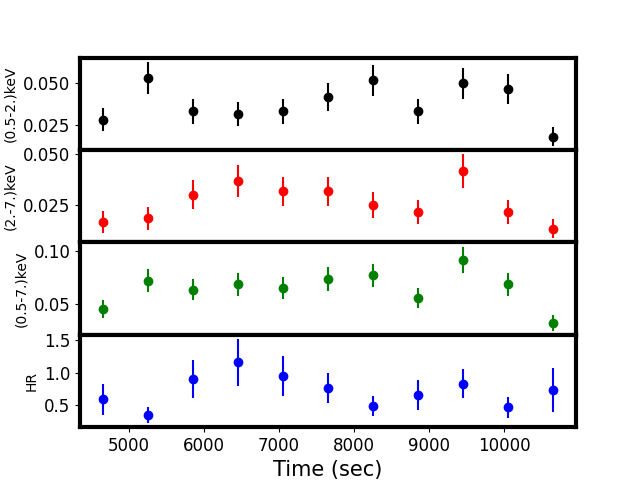}
\hspace*{-0.85cm}\includegraphics[scale=0.45]{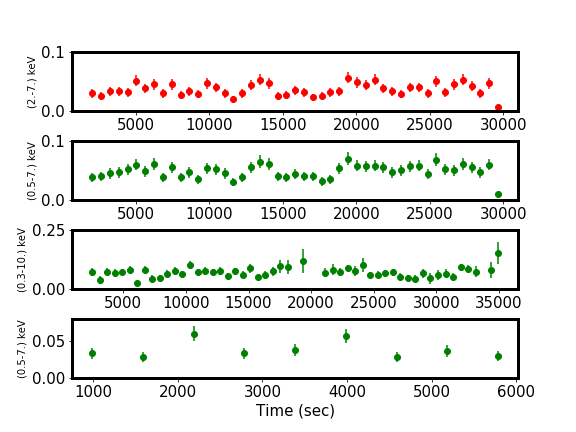}
         }

    \caption{Left panel: Light curves with a binning of 600 sec for the object J0131+5545 from {\it Chandra} observations. 
Here, from the top, the flux variations in the energy ranges of 0.5$-$7 keV, 2$-$7 keV and 0.5$-$7 keV are shown. The bottom panel shows the variation of HR. Right Panel: Top two panels for probable variable J1347+1217 from {\it Chandra} observations with the flux variations in the energy range of 2$-$7 keV (hard band) and 0.5$-$7 keV (total band). The second middle and bottom panels are for possible variables J1939$-$6342 and J1723$-$6500 from {\it XMM-Newton} and {\it Chandra} observations, respectively, in the total band.}
\label{figure-2}
\end{figure*}

The uncertainty in F$_{var}$ is defined as \citep{2003MNRAS.345.1271V}
\begin{equation}
F_{var,err} = \sqrt{\left( \sqrt{\dfrac{1}{2n}}\dfrac{\bar\sigma^2_{err}} {\bar{x}^2 F_{var}}\right)^2 +  \left(\sqrt{\dfrac{\bar\sigma^2_{err}}{n}} \dfrac{1}{\bar{x}}\right)^2}
\end{equation}
In this work, we have considered the entire light curve as one interval to calculate the variability amplitude \citep{2003MNRAS.345.1271V, ponti2012}. This is due to the fact that the net exposure time is relatively small ($<$50 ksec) for most of the sources and even in sources with longer exposure, no significant variability was identified.
We calculated F$_{var}$ in the soft (0.3$-$2 keV), hard (2$-$10 keV) and the total (0.3$-$10 keV) energy bands. For {\it Chandra}, these were selected as 0.5$-$2 keV, 2$-$7 keV, and 0.5$-$7 keV as the soft, hard, and total bands. The F$_{var}$ was computed only in the single 3$-$20 keV band for the {\it NuSTAR} data. We considered a source to have shown 
flux variations only if $\sigma^2_{NXV}$ > 0 and F$_{var}$ $>$ 3 $\times$ $F_{var,err}$ in any of the bands. Those which satisfy the criterion
2 $\times$ $F_{var,err}$ $<$ F$_{var}$ $<$ 3 $\times$ $F_{var,err}$ in any of the bands were designated as possible variable.
The observations with no entries in fractional variability did not satisfy the $\sigma^2_{NXV}$ criterion. 
The results of the variability analyses from 
{\it XMM-Newton}, {\it Chandra} and {\it NuSTAR} observations are presented in Table \ref{table-2}.
The light curves of the sources found to be variable or probable variable in all three bands or in any of the bands from {\it XMM-Newton} and {\it Chandra} observations are shown in Figs. \ref{figure-1} and \ref{figure-2} respectively. The light curves for the sources found to be probable variable in only one band or two bands in either {\it Chandra} or {\it XMM-Newton} observations are shown in the right panel of Fig. \ref{figure-2}. 

Of the thirteen sources with {\it XMM-Newton} observations, F$_{var}$ could be estimated for only six sources and $\sigma^2_{NXV}$ was negative for the remaining sources.
Using the above criteria, we found unambiguous evidence of flux variations in two sources, namely J0713+4349 and J1326$+$3154, in 
all three energy bands. The source J1939$-$6342 was found to be a probable variable in the total band as 2 $\times$ $F_{var,err}$ $<$ F$_{var}$ $<$ 3 $\times$ $F_{var,err}$.

Similarly, of the 8 sources, with {\it Chandra} observations, F$_{var}$ could be estimated for seven sources. Out of these, one source, namely, J0131+5545, was found to be variable in total energy band but a probable variable in the soft and hard bands. J1347+1217 was found to be a possible variable in the hard and total energy bands, while J1723$-$6500 was found to be a probable variable only in the total band.
Among the 6 sources with {\it NuSTAR}
observations, F$_{var}$ could be estimated for two sources, and none of these were found to be variable in the 3$-$20 keV band.

Within errors, the flux variations
in all the bands are found to be similar in J0713+4349. In the case of 
J1326$+$3154, F$_{var}$ is larger in the hard and total energy bands than in the soft band. This finding indicates a possible dominance of the variable jet emission in the hard X-ray band. However, it can also be due to variation in the seed photon flux in the disk-corona interaction \citep[cf.][]{2005MNRAS.363.1349G}.

\begin{table}
\centering
\caption{Results of the variability analysis for CSOs with 
$\sigma^2_{NXV}$ > 0 in at least one energy band. The last column shows whether the source is variable or not. 'Y' stands for yes if it is variable in any of the bands, while 'N' indicates that it is not variable in any of the bands.}
\label{table-2}
\begin{tabular}{llllc}
\hline 
Name       &  \multicolumn{3}{c}{F$_{var}$ $\pm$ F$_{var,err}$}&Variable\\
\hline 
{\it XMM-Newton} &0.3$-$2 keV     & 2$-$10 keV        & 0.3$-$10 keV &\\ 
J0713+4349 & 0.18$\pm$0.05  & 0.28$\pm$0.09 & 0.21$\pm$0.04 & Y\\ 
J1220+2916 & 0.01$\pm$0.01  &         -      &       -     & N   \\
J1326+3154 & 0.70$\pm$0.07  & 1.17$\pm$0.27 & 0.88$\pm$0.09& Y \\ 
J1407+2827 & 0.05$\pm$0.15  &       -        &     -       & N   \\ 
J1723$-$6500 &          -      & 0.04$\pm$0.25 & 0.04$\pm$0.07 & N\\ 
J1939$-$6342 & 0.08$\pm$0.09  &     -          & 0.15$\pm$0.06$^a$ & PV\\  
\hline 
{\it Chandra} &   0.5$-$2 keV       &    2$-$7 keV   &     0.5$-$7 keV         &\\
J0131+5545 & 0.20$\pm$0.08$^a$ & 0.23$\pm$0.10$^a$ & 0.19$\pm$0.06 &Y \\
J0405+3803 & - & 0.44$\pm$0.30 & - &N\\
J0713+4349 & 0.05$\pm$0.11 & 0.02$\pm$0.17 & 0.08$\pm$0.06 &N\\
J1220+2916 & - & 0.11$\pm$0.10 & - & N\\
J1347+1217 & 0.07$\pm$0.21 & 0.19$\pm$0.08$^a$ & 0.14$\pm$0.06$^a$ & PV \\
J1407+2827 &  -&0.08+0.09  & 0.12+0.08  &N\\
J1723$-$6500 & 0.20$\pm$0.11 & 0.12$\pm$0.20 & 0.23$\pm$0.09$^a$ &PV \\
\hline
 {\it NuSTAR} &   \multicolumn{2}{c}{3$-$20 keV}      &  &         \\
   & FPMA & FPMB & &\\
J2327+0846 & 0.08$\pm$0.11 &- & & N \\
J1407+2827 & 0.15$\pm$0.09 & - & & N\\
\hline
\end{tabular}
$^a$: possible variable (PV)
\end{table} 

\subsection{Spectral variability}
\begin{figure*}

      \includegraphics[scale=0.45]{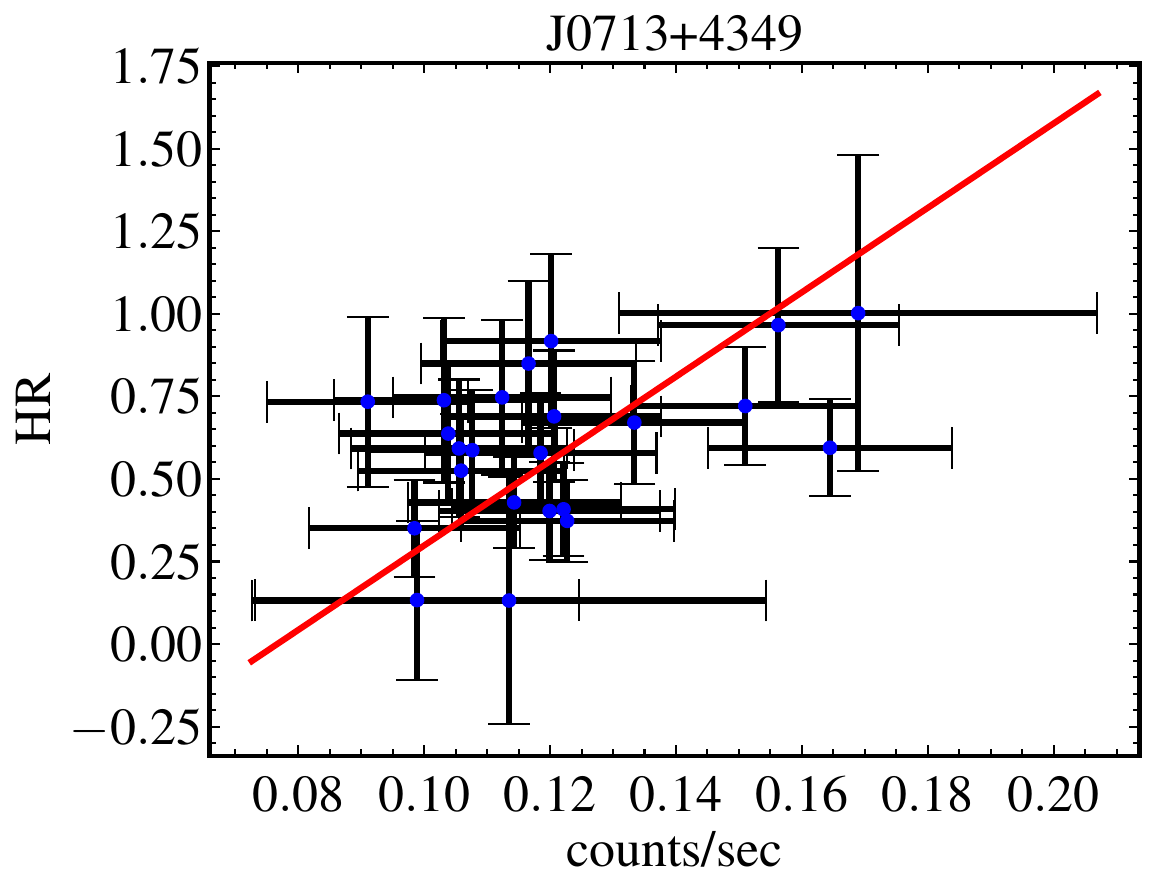}
      \includegraphics[scale=0.45]{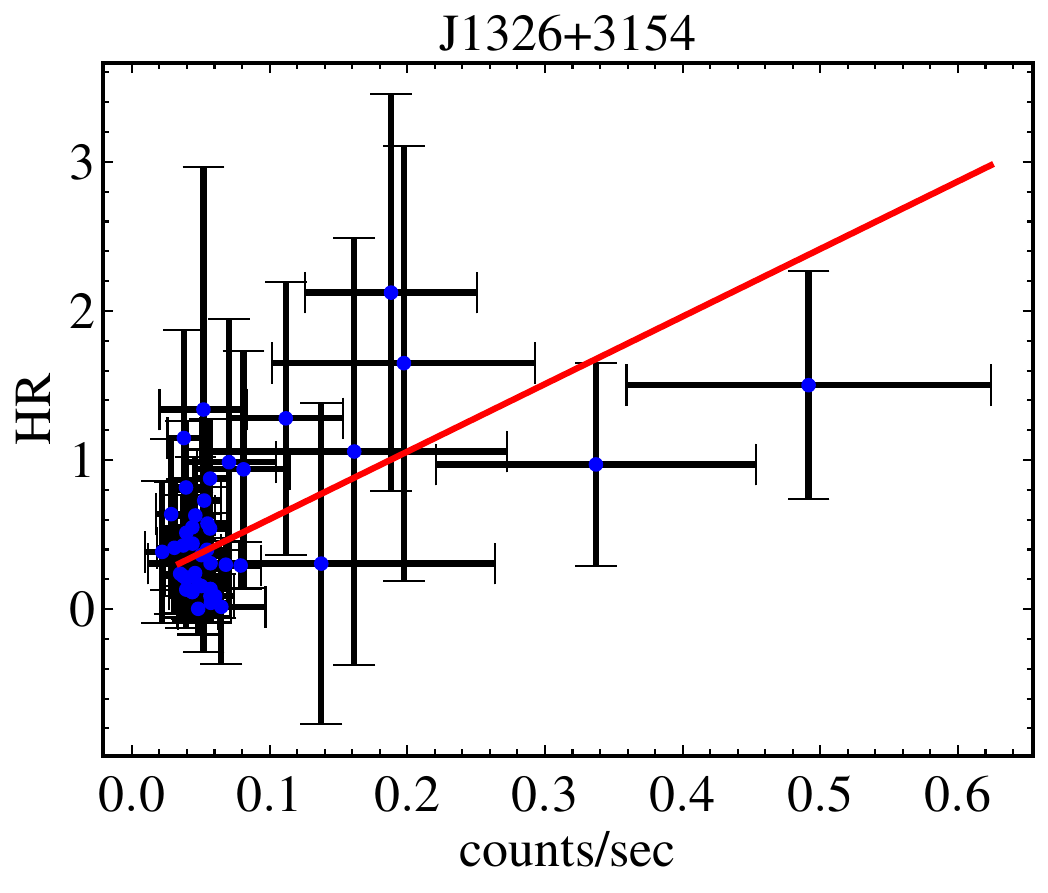}

    \caption{HR plotted as a function of count rate in the 0.3$-$10 keV energy 
range for the two sources that showed spectral variations in {\it XMM-Newton} observations. The names of the objects are given in each panel. The red solid line is the 
weighted linear least-squares fit to the data.}
    \label{figure-4}
\end{figure*}

To investigate the spectral variations in the sources that have shown
flux variations in a model-independent way, we calculated the 
hardness ratio (HR) and then examined the variation of HR with their
total count rate (CR). We define the HR as
\begin{equation}
HR = \frac{H}{S}
\end{equation}
Here, $H$ and $S$ are the count rates in the hard and soft bands, respectively. The variation of HR with the count rate in 
the total band are given in Figure \ref{figure-4}. 
To quantify the significance of the correlation between HR and 
the count rate in the total band, we fitted the observed 
points in the HR versus count rate diagram using a linear
function of the form HR = a $\times$ CR + b, where a is the slope 
and b is the intercept. During the fit, we took into account the errors in 
both HR and count rate. The results of the fits are given in Table \ref{table-3}. 
We found a harder when brighter (HWB) trend in two sources, namely J0713+4349 and J1326+3154, that are variable as per our criterion. The observations of these two sources are from {\it XMM-Newton}.
The variation of HR with the total count rate for the two sources are shown in 
Figure \ref{figure-4}. Also, shown in the same figure are the linear least-square fits to the data. 
No significant trend was seen in the {\it Chandra} observations of J0131+5545. 

\begin{table}
\centering
\caption{Results of correlation analysis. Here, R and p are the correlation coefficient and the probability of no correlation, respectively.}
\label{table-3}
\begin{tabular}{lrrrl}
\hline 
Name  & Slope & Intercept &  R & p \\ 
\hline 
{\it XMM-Newton} &         &                 &      &           \\
J0713+4349 & 9.24$\pm$2.88 &  $-$0.53$\pm$0.34 &  0.44 & 0.03 \\
J1326+3154 & 4.53$\pm$1.53 & 0.15$\pm$0.11 &  0.56& 5.75$\times$10$^{-5}$\\ 
\hline
{\it Chandra}    &         &                     &        &       \\
J0131+5545 &  0.92$\pm$0.38 & $-$4.75$\pm$5.33  & 0.04 & 0.91 \\
\hline 
\end{tabular}
\end{table} 

\subsection{Spectral analysis}
\begin{figure}
\centering
   \includegraphics[scale=0.45]{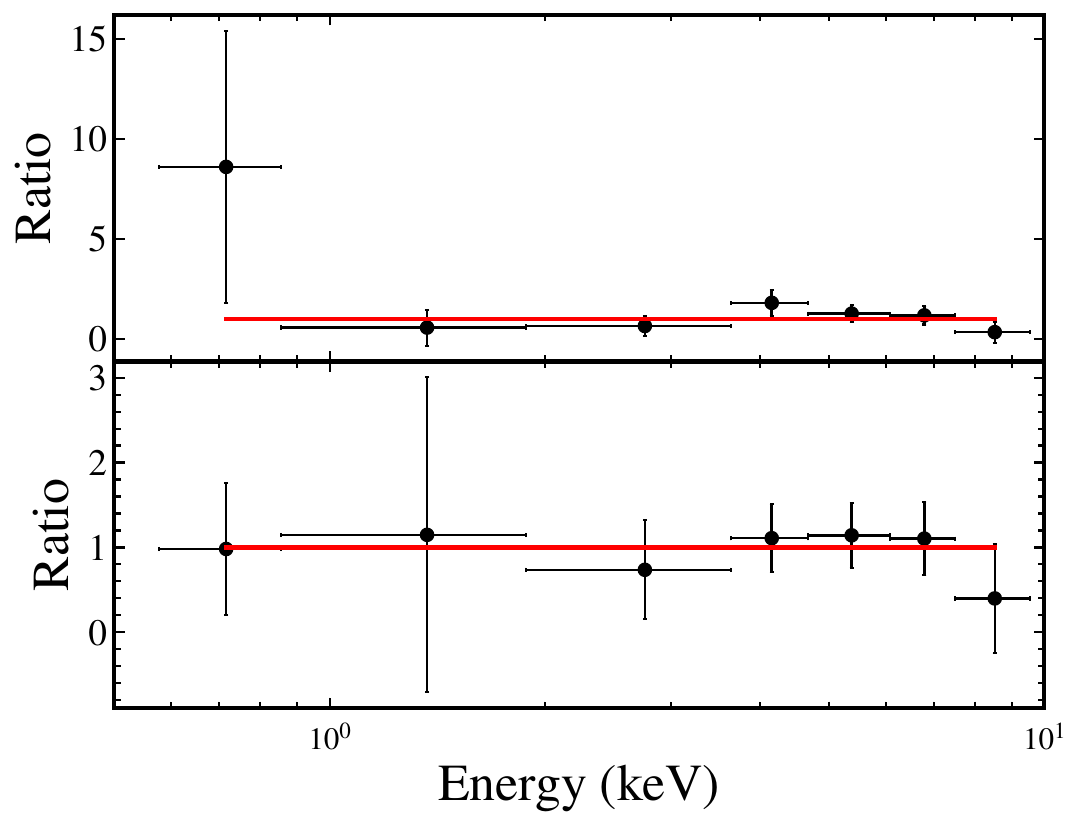}
\caption{Ratio of {\it XMM-Newton} data to model for CSO J0111+3906. Top:  power-law modified with Galactic and intrinsic absorbing columns (Model A). Bottom: model-A is modified with a blackbody component (model-B). The red line is a constant at ratio = 1.}
\label{j0111}
\end{figure}

In our sample, 13 sources have observations from {\it XMM-Newton}, 8 sources have observations from {\it Chandra} and 6 sources have observations from {\it NuSTAR}. The spectral analysis for all CSOs have been reported in the literature. However, to gain a more consistent understanding of their nature, we reanalyzed the archival data to determine key parameters such as the photon index ($\Gamma$), hard X-ray luminosity (L$_{2-10~\rm keV}$), and Eddington ratio ($\lambda_{Edd}$). Hence, we carried out a phenomenological model fits to the observed spectra using XSPEC V12.9 \citep{1996ASPC..101...17A} using powerlaw for primary X-ray emission, blackbody or diffuse emission model for soft X-ray excess and reflection model for reprocessed emission. For this, we fitted the spectra with models that have the following form in XSPEC
\begin{equation}
TBABS(APEC/BBODY + ZTBABS*POWERLAW + GAUSS)
\end{equation}
where {\sc apec} (blackbody) model is to account for the diffuse emission (thermal emission) and {\sc gauss} model is for line emission as reflection features. We used {\sc tbabs} to model the Milky Way Galactic hydrogen column density ($N_{H,Gal}$), the values of
which were taken from \cite{2013MNRAS.431..394W} and were fixed at the respective values and {\sc ztbabs} to model the intrinsic column density ($N_{\rm H,z}$).
First, we fitted the spectra of all the sources with a model consisting of an absorbed power-law as `model A' ({\sc{tbabs * ztbabs * powerlaw}}) with the normalization and the 
photon index ($\Gamma$)\footnote{The photon spectra of AGN are represented as a power-law with the form $N_E \propto E^{-\Gamma}$ (photons cm$^{-2}$ s$^{-1}$ keV$^{-1}$)} and intrinsic column density ($N_{\rm H,z}$) as free parameters. Depending on the residual in the spectra, 
we fitted the spectra of those sources with equation 6 as `model B'.
Then, motivated by the X-ray modeling of a few CSOs \citep{Sobolewska_2023, sobo2019} that have hard photon index, we explored the complex model for a toroidal reprocessor to account for the residuals in the hard band. The models of our choice are i) {\sc pexrav}: model that accounts power-law spectrum reprocessed by neutral material \citep{Magdziarz1995} accounting for a fraction of X-ray continuum reflected by distant material into the line of sight. ii) {\sc xillver}: model that accounts for related X-ray reflection emission from the illuminated accretion disk \citep{Garcia_2013} iii) {\sc borus}: model that considers the reflection of the primary X-ray continuum > 4 keV from cold matter, presumably a toroidal structure around the central black hole \citep{Balokovic_2018}. We studied four sources, J1407+2827, J1511+0518, J2022+6136 and J2327+0846, using combined data from {\it XMM-Newton} and {\it NuSTAR} observations. For the first three sources, we used a more detailed reflection model called {\sc borus} to analyze their X-ray properties, while for the source J2327+0846, we used the reflection model {\sc xillver}. In all the model fits, the errors in the parameters were calculated at 90\% confidence ($\chi^2$ = 2.71 criterion). The results on the individual sources are described in the Appendix, except for two CSOs, i.e. J0111+3906 and J1945+7055. For these two CSOs, the thermal component has not been explored earlier in the literature. The spectral properties of these two objects, J0111+3906 and J1945+7055, are discussed below. The obtained spectral fitting parameters for all are consistent with those reported in the literature. The spectral parameters are given in Tables \ref{table-5} and \ref{table-4} for CSOs using archival {\it XMM-Newton}, {\it Chandra} and {\it NuSTAR} observations. We refer to the Table \ref{tab:tabcps} for hard X-ray luminosity L$_{2-10}$ and Eddington ratio ($\lambda_{Edd}$) based on these derived spectral parameters.
Our results are described in the
following sections.\\
\begin{center}
\text{J0111+3906:}
\end{center}
Previous studies revealed that this source often exhibit significant intrinsic absorption, with \cite{vink2006} finding an intrinsic column density of N$_{H,z}$ = 57$\pm$20 $\times$ 10$^{22}$ cm$^{-2}$ using an absorbed power-law model. In their analysis, they froze the photon index ($\Gamma$) at 1.75, suggesting that the source is heavily absorbed by a dusty torus. However, they did not account for residuals in the soft X-ray band or investigate flux variations in the source, leaving some uncertainties in the nature of the absorption and the spectral characteristics. In our reanalysis using archival data, we first applied Model A (absorbed power-law) but obtained a $\chi^2=1.72$ with unconstrained value of $\Gamma$ and N$_{H,z}$ as 2.46$^{+3.69}_{-3.16}$ and 87.38$^{+188.9}_{-88.19}$ $\times$ 10$^{22}$ cm$^{-2}$ respectively. Therefore, we froze $\Gamma$ to 1.80 as the data are statistically limited. We obtained a slightly better fit with $\chi^2$=1.47 and N$_{H,z}=61.84^{+45.62}_{-30.24} \times 10^{22}$ cm$^{-2}$. The residuals in the soft band are shown in Figure \ref{j0111} (top panel). To examine the further improvement in the fit, we employed Model B, which included a thermal component along with the power-law component. We found a thermal emission kT of 0.11$^{+0.03}_{-0.02}$ keV with $\chi^2$ = 0.45 (Figure \ref{j0111}). 
From this analysis, we determined $N_{\rm H,z} = 66.7^{+47.5}_{-31.8} \times 10^{22}$ cm$^{-2}$, consistent with previous findings \citep{vink2006}. This source has been studied for X-ray flux variations for the first time. 
However, we found the source to remain non-variable during the period of the observation. \\


\begin{figure}
\centering
   \includegraphics[scale=0.45]{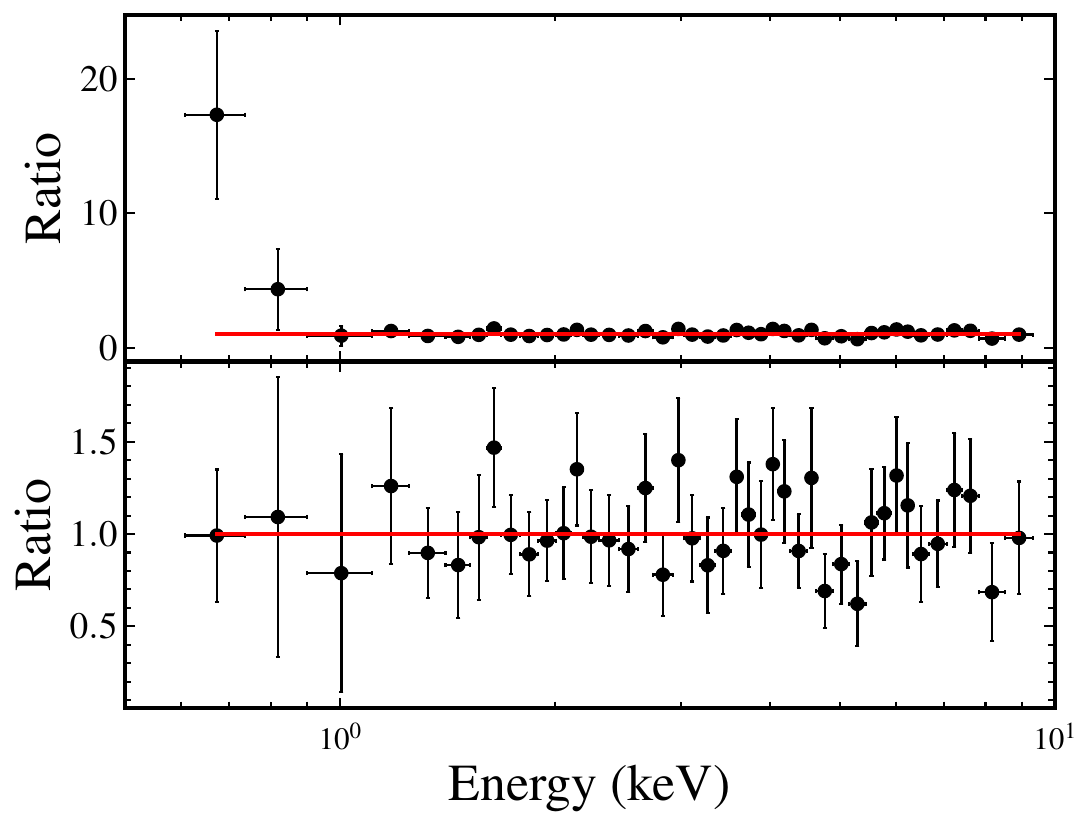}
\caption{Ratio of {\it XMM-Newton} data to model for CSO J1945+7055. Top: power-law modified with Galactic and intrinsic absorbing columns (Model A). Bottom: model-A is modified with a blackbody component (model-B). The red line is a constant at ratio = 1.}
\label{j1945}
\end{figure}

\begin{center}
\text{J1945+7055:}
\end{center}
{\it BeppoSAX} observation of this source revealed a possible Compton thick nature with $N_{\rm H,z} > 2.5 \times 10^{24}$ cm$^{-2}$ \citep{2003A&A...401..895R}.
Analysis of the {\it Chandra}
data of the same source by \cite{Siemiginowska2016} constrained $\Gamma$ = 1.7$\pm$0.4 and $N_{\rm H,z} = 1.7^{+0.5}_{-0.4} \times 10^{22}$ cm$^{-2}$, suggesting it to be not a Compton thick source.
Moreover, they obtained an upper limit of 7.2 keV for the equivalent width of 
the Fe K$\alpha$ line. Based on {\it XMM-Newton} data analysis by \cite{sobo2019}, J1945+7055 is a mild obscured source with an equivalent width of 
the Fe K$\alpha$ line $<$ 0.2 keV. 
In this work, we re-analysed the data acquired by 
{\it XMM-Newton} on 21 October 2016. Based on the absorbed power-law model 
fit to the spectrum, we obtained $\Gamma$ = 1.13$^{+0.27}_{-0.25}$ and $N_{\rm H,z} = 1.55^{+0.60}_{-0.50} \times 10^{22}$ cm$^{-2}$. Adding a thermal component to the absorbed power-law model, we obtained 
kT = 0.18$^{+0.01}_{-0.01}$ keV and blackbody normalization of 8.26$_{-7.59}$ $\times$ 10$^{-7}$ cm$^{-5}$. Since the normalization parameter could not be well constrained, we froze it to the above-mentioned value. 
The ratio of {\it XMM-Newton} data to model-A and model-B are shown in Figure \ref{j1945}. 
Considering errors, the results from the spectral fits to {\it XMM-Newton} data are in agreement with that obtained by \cite{Siemiginowska2016} from {\it Chandra} data. The values of $\Gamma$ and N$_{H,z}$ are also in agreement with the results of \cite{sobo2019}. 

\section{Discussion}
The aim of this work is to investigate the X-ray timing and spectral properties
of a bona fide sample of CSOs. Two sources, J0713+4349 and J1326+3154, were found to show
evidence of flux variations from {\it XMM-Newton} observations in all three bands, while from {\it Chandra} observations one source, J0131+5545, 
showed variability in one of the bands. For most of the sources, 
a simple power-law modified by absorption is an adequate representation of 
the observed spectra, although there are exceptions. 
The results of X-ray flux variations for the sources J0131+5545, J0713+4349 and J1326+3154 are being reported for the first time. However, the spectral analysis results for most sources are already available in the literature. Nevertheless, we have re-analysed the X-ray spectra of the bona fide sample in a homogeneous manner and compared with the results in the literature. 

\subsection{Flux and Spectral variability}
 
Although, the X-ray variability behaviour of the misaligned jetted AGN population is not well understood, the X-ray emission observed from FR I and II radio galaxies is thought to be produced due to inverse Compton process \citep[e.g.,][]{Torresi:2012rp,2023PASJ...75.1124M,krol_2024}. This includes the upscattering of the synchrotron photons and/or photons produced by the accretion disk, BLR, or torus, by the same electron population. Therefore, the detection of rapid flux variations ($\sim$hours timescale or shorter) indicates a small emission region size close to the base of the jet. On the other hand, the seed photons for the inverse Compton mechanism can also be provided by radio lobes and/or cosmic microwave background. Since the radiative energy density of these components dominate far from the central black hole, fast variability is not expected due to much larger emission region, assuming the jet to be of conical shape. Additionally, the X-ray emission can also be produced by the shocked interstellar medium \citep{ODea_2000}. Considering the radio-quiet AGN population, X-ray flux variations of the considered timescales are often connected with the disk-corona interaction, i.e., regions near the central black hole \citep[see, e.g.,][]{2021MNRAS.508.1798P}. Only a handful of very bright ($F_{\rm X}>10^{-11}$ erg cm$^{-2}$ s$^{-1}$) misaligned radio-loud AGN have exhibited similar observational features \citep[][]{Gliozzi_2009,Lohfink_2013,2017ApJ...841...80L}.

Considering CSOs, we tried to quantify the extent of short timescale of flux variations by calculating the flux doubling/halving timescales \citep[see, e.g.,][]{2013A&A...555A.138F}. However, no significant rapid flux variations were found. The lack of rapid X-ray flux variability could be due to the fact that the high-energy emission in CSOs can be produced via inverse Compton scattering of the low-energy photons by the electrons present in the expanding radio lobes \citep[cf.][]{2008ApJ...680..911S}. 
Furthermore, three CSOs, namely, J0713$+$4349, J1326$+$3154 and J1347+1217, are among the brightest sources compared to other CSOs, suggesting X-ray brightness can be related to their X-ray variability. Measuring the flux variations for the faint sources could be challenging. 

Although a majority of the sources in our sample do not show any correlation between HR and count rate, two sources, J0713$+$4349 and J1326$+$3154, do exhibit a positive correlation, i.e., a harder when brighter trend. The HWB trend shown in these CSOs has not been reported earlier. These two sources
appear similar to the class of accreting black holes such as black hole X-ray binaries (BHXRBs) where the CSOs may be the analogues of the hard counterpart of BHXRBs below $\sim$1\% Eddington luminosity \citep{Wu2008, gu2009, Sobolewska2011a, trichas2013, Emmanoulopoulos2012}. Also, a similar trend has been observed in other types of AGN such as low luminosity AGN, LINERS, and high synchrotron peaked (HSP) blazars \citep{priyanka2016, Moravec2022, Fern2023A}. 
This trend seen in the sources studied here might be attributed to the emergence of a hard X-ray tail produced in the jet.\\

The correlation between radio and X-ray luminosity of both CSOs and X-ray binaries links the presence of radio jets to the properties of the accretion flow in these accreting black holes. CSOs are characterized by $\Gamma\sim$1.1-2.5, $\lambda_{Edd}\sim$ 0.0001-0.15 and thermal plasma/blackbody temperature kT$\sim$0.11-1.09 keV. Hence, the main physical processes known for power-law X-ray emission, as in the case of the majority of CSOs and binary black holes \citep{MOTTA2021101618}, are IC scattering from the disc/corona geometry like in an AGN or IC scatterings through relativistic jets. But a broadband radio to $\gamma$-ray spectral energy distribution will help to understand better the physical process taking place in the nuclear regions \citep[e.g.,][]{2022ApJ...941...52S}.

\subsection{Origin of X-ray emission}
X-ray emission is ubiquitous in AGN. However, the mechanisms by which X-rays are produced differ among different categories of AGN \citep{2017FrASS...4...35P}. In the radio-quiet category of AGN, X-rays are believed to be produced in a region called the corona, consisting of hot electrons and situated close to the vicinity of the central supermassive black hole. The electrons in the corona, inverse Compton scatter the optical/UV accretion disk photons to X-rays. This naturally implies a connection between X-ray corona and the accretion disk in AGN \citep{1993ApJ...413..507H}. In the radio-loud category of AGN, the observed X-rays are predominantly from the relativistic jet with a negligible contribution from the X-ray corona for beamed AGN \citep{2022Galax..10....6F}. For non-beamed jetted AGN, an additional significant contribution from X-ray corona is possible and has been observed in a few bright radio galaxies \citep[][]{Lohfink_2013,2017ApJ...841...80L}.

For most of the sources studied in this work, we found $\Gamma$ to be in the range between 1.5 and 2.5. These values of $\Gamma$ are similar to that known for the Seyfert type AGN, where the X-ray
emission is predominantly from the disk/corona. This may be the case, particularly for J0713+4349, J1347+1217, 
J1407+2827, J1511+0518 and J2022+6136 where the FeK$\alpha$ line is detected. Such detection above the continuum points to the case where the accretion power dominates the jet as seen in 3C 273 \citep{Paola_2004}. Another possibility of observed iron lines in CSOs may be produced by a reflection of the lobes continuum from the surrounding cold dust \citep{krol_2024}.
From an analysis of a large number of sources, \citet{Liao2020}, using the
linear relation between the luminosity in the radio band at 5 GHz and the
luminosity in the X-ray band in 2$-$10 keV, conclude that the
X-ray emissions from compact objects are from the jet and inconsistent with
the theoretical prediction of accretion flows as the cause of X-ray emission. Among the sources that are variable, two showed spectral variability with a HWB trend similar to that observed in other accreting sources. This could be attributed to the dominance of the jet emission to X-rays and can be explained by synchrotron self Compton models \citep{Krawcz2004,Emmanoulopoulos2012}.

\begin{table*}
	\centering
	\caption[]{The list of CSOs with measured X-ray luminosity and Eddington ratio. Columns show (1) J2000 name, (2) Black hole mass (M$_{BH}$), (3),(5) References, (4) Radio luminosity at 5 GHz (L$_{5GHz}$) (6) absorption corrected rest-frame X-ray luminosity in 2-10. keV band (L$_{X(2-10)}$) (7) photon index (8) Eddington ratio ($\lambda_{Edd}$). (References. (1) References in \cite{Liao2020}, (2) \cite{Tremblay2016} (3) \cite{tengstrand2009} , (4)\cite{an2012}, (5) \cite{Wojtowicz_2020}, (6) \cite{Fan2016}, (7) \cite{Woo_2002}, (8) \cite{willett2010}. Parameters listed in columns (6-8) were obtained in this work. ‘f’ indicates that the values are frozen to the values given in the table while fitting the spectra.}
\begin{tabular}{lllllllll}
\hline
\hline 
 Source  &  M$_{BH}$ & Ref & logL$_{5GHz}$ & Ref & logL$_{X(2-10)}$  &   $\Gamma$ & log $\lambda_{Edd}$ \\
 & $\times$10$^{8}$ M$_{\odot}$ & & erg s$^{-1}$ & &  erg s$^{-1}$  & &  \\
 (1) & (2) & (3) & (4) & (5) & (6) & (7) & (8)  \\

 \hline
J0029+3456& 3.7 & (1) &43.65&(1)&44.13$^{+0.03}_{-0.04}$&1.61$^{+0.24}_{-0.27}$&	-1.29	 \\
J0111+3906 & 0.79& (1)&43.87&(1)&43.62$^{+0.14}_{-0.22}$&1.80 (f)&	 -1.22 	\\
J0713+4349 & 2.5 & (6)&43.76&(1)&44.60$^{+0.04}_{-0.05}$ 	&1.52$^{+0.14}_{-0.13}$ &-0.82\\
J1148+5924 & 20&(8)	&39.44	&(1)&40.75$^{+0.07}_{-0.07}$&1.75$^{+0.35}_{-0.35}$& -4.04  \\
J1220+2916 & -& - & 38.58&(2)&40.60$^{+0.01}_{-0.01}$ &2.07$^{+0.03}_{-0.03}$ &-	\\
J1326+3154 & 15.85 &(6)	&43.60&(1)&43.62$^{+0.06}_{-0.07}$&	1.65$^{+0.22}_{-0.19}$ &	-2.44\\
J1347+1217 & 1.99&(6)	&42.70&(3)&	43.71$^{+0.03}_{-0.03}$&1.63$^{+0.31}_{-0.28}$&-1.17\\
J1407+2827 & 5.31& (1)&42.78&(3)&42.62$^{+0.06}_{-0.07}$ &1.43$^{+0.15}$ &		-2.69\\
J1511+0518 & 4 &	(5)&41.69&(1)&42.57$^{+0.42}_{-0.38}$&	1.62$^{+0.27}$	&-2.96\\
J1609+2641& 3.98 & (5) &43.65&(4)&	43.31$^{+0.06}_{-0.06}$ &	2.27(f)&	-2.22\\
J1723-6500 & 3.16 & (5)& 41.06&(4)&41.24$^{+0.02}_{-0.02}$&	1.80$^{+0.09}_{-0.09}$	&-4.20\\
J1939+6342 & 3.16 &(5)& 43.34&(4)&43.09$^{+0.05}_{-0.06}$&		1.68$^{+0.1}_{-0.18}$	&-2.39\\
J1945+7055 & 3.16 & (5)&41.87	&(4)&43.08$^{+0.04}_{-0.04}$&1.17$^{+0.27}_{-0.24}$&-2.40	\\
J2022+6136 & 7.94 &(5)	&43.24&(4)&43.84$^{+0.05}_{-0.06}$&	1.51$^{+0.09}_{-0.07}$&		-1.69 \\
J2327+0846 & 3.63 &(7)	&42.70&(3)&42.02$^{+0.18}_{-0.06}$&	1.63$^{+0.18}_{-0.03}$&		-1.30\\

\hline
\hline 
\end{tabular}
\label{tab:tabcps}
\end{table*}

\begin{figure*}
\hbox{
\includegraphics[scale=0.45]{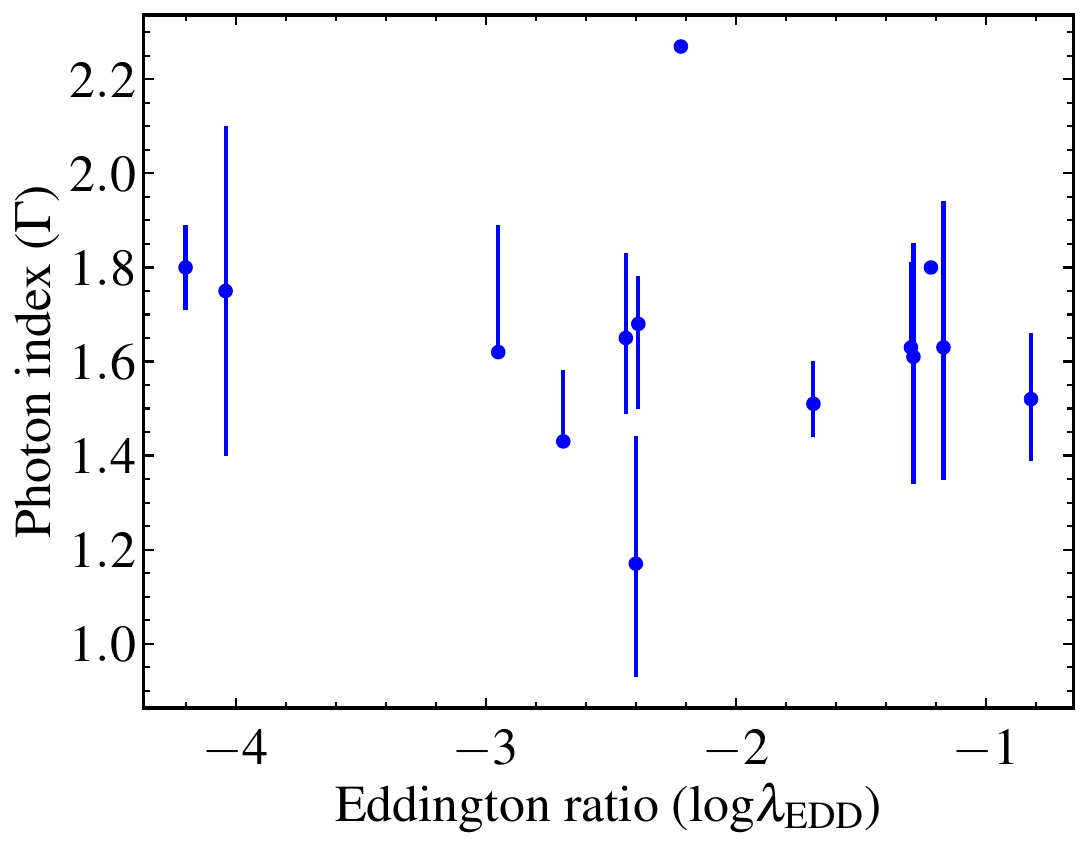}
\hspace*{1.0cm}\includegraphics[scale=0.45]{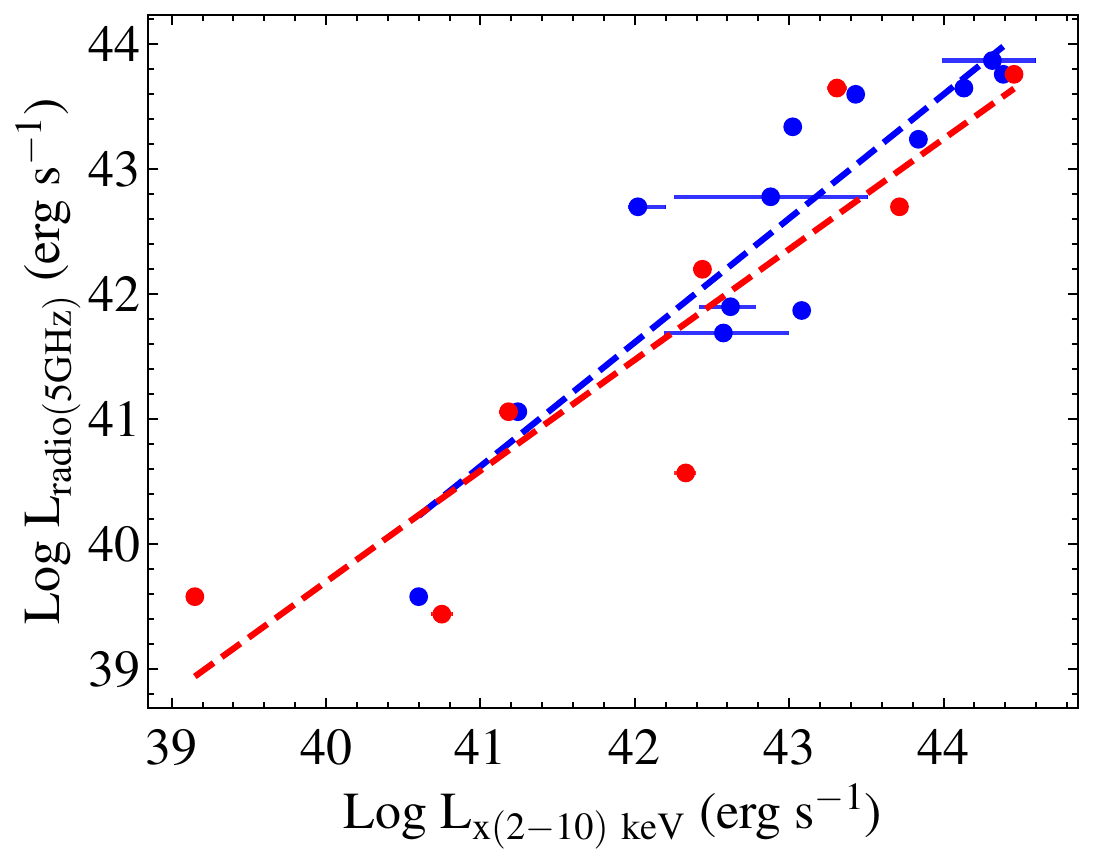}
    }
\caption{Left panel: The plot of $\Gamma$ versus $\lambda_{Edd}$ from both {\it XMM-Newton} and {\it Chandra} data. The filled circles with no error bars denoted fixed $\Gamma$ values. Right 
panel: Correlation between X-ray and radio luminosity. Here the blue colour refers to {\it XMM-Netwon} observations and the red
colour refers to {\it Chandra} observations. The solid and dashed lines in
both the figures are the linear least squares fit to the data. }
\label{figure-7}
\end{figure*}

Studies available in the literature point to a positive correlation between the X-ray photon index and the accretion rate parameterized by the Eddington ratio above a critical Eddington ratio of 0.01 \citep{Wu2008, gu2009, Sobolewska2011a, trichas2013, Emmanoulopoulos2012, 2013MNRAS.433.2485B} ($\lambda_{Edd} = L_{Bol}/L_{Edd}$, where, $L_{bol}$ is the bolometric
luminosity and $L_{Edd}$ is the Eddington 
luminosity).
To explore key disk parameter which influences conditions in the corona, \cite{2013MNRAS.433.2485B} considered a sample of radio-quiet AGN and reported a strong correlation between $\Gamma$ and $\lambda_{Edd}$. 
From detailed studies over a large range of $\lambda_{Edd}$, the slope of the relationship depends on the range of $\lambda_{Edd}$, being positive at $\lambda_{Edd}$ > 0.02 \cite[e.g.][]{2015MNRAS.447.1692Y}. Such a positive 
correlation between $\Gamma$ and $\lambda_{Edd}$ is suggestive of a 
connection between the accretion disk and the X-ray corona. For our CSOs, $\lambda_{Edd}$ values are in the range of 10$^{-4}$ - 10$^{-1}$.
In our study, $L_{bol}$ is calculated as $L_{bol}$ = K$_X$ $\times$ $L_{X \rm (2-10~keV)}$ where the bolometric correction K$_X$ is calculated using Equation 3 from \cite{Duras2020}. To determine K$_X$, we used the parameters a = 15.33, b = 11.48, and c = 16.20, which are generally adopted for both Type 1 and Type 2 AGN. We found K$_X$ to range between 15.34 and 23.17. 
The Eddington ratio estimates for a few CSOs in our sample are consistent with those reported by \cite{Wojtowicz_2020}. However, in \citet{Wojtowicz_2020}, the bolometric luminosity was derived using a heterogeneous method based on the H$_{\beta}$/[OIII] or 12-micron emission line. In contrast, we have calculated the bolometric luminosity uniformly for all sources by applying the X-ray bolometric correction in the 2-10 keV energy range \citep{Duras2020}. For five CSOs, the Eddington ratios differ from those in \citet{Wojtowicz_2020}. This discrepancy may be due to differences in the adopted methods to compute $L_{bol}$. 
 The correlation between $\Gamma$ and $\lambda_{Edd}$ is shown in Figure \ref{figure-7} for the CSOs in our sample having black hole mass measurements. Linear least-squares fit to the data yield a correlation coefficient of 0.14 with a probability of no correlation $p$ of 0.63. We thus 
found no correlation between $\Gamma$ and $\lambda_{Edd}$.

Furthermore, we show the correlation between the radio 
luminosity at 5 GHz ($L_{5 GHz}$) and the intrinsic X-ray luminosity in the 
2$-$10 keV band ($L_{2-10 keV}$) in the right panel of Figure \ref{figure-7}. Considering sources only with {\it Chandra} observations, from linear least-squares fit to the data, we find a linear correlation coefficient of 0.90 and a probability of no correlation of 0.003. Similarly, for sources with {\it XMM-Newton} observations, we find a linear correlation coefficient of 0.90 and
a probability of no correlation of 2.53 $\times$ 10$^{-5}$. Therefore, a positive correlation is observed. The positive correlation between $L_{5 GHz}$ - $L_{2-10 keV}$ and no-correlation
between $\Gamma$ and $\lambda_{Edd}$ show that for the sources plotted in 
Fig. \ref{figure-7}, the observed X-ray emission could be due to processes in the
relativistic jets of CSOs. However, we note that the sources J0713+4349, J1347+1217, 
J1407+2827, J1511+0518 and J2022+6136 also follow the correlation between $L_{5 GHz}$ and $L_{2-10 keV}$. The X-ray spectra of these objects exhibit a prominent Fe K$\alpha$ line. Therefore, it is likely that the X-ray emission in 
these sources may have a significant contribution from the accretion disk-jet coupling due to the emission of a radiatively inefficient accretion flow \citep[e.g.,][]{merloni2003}, indicating a complex radiative environment. 

\begin{figure}
\includegraphics[scale=0.45]{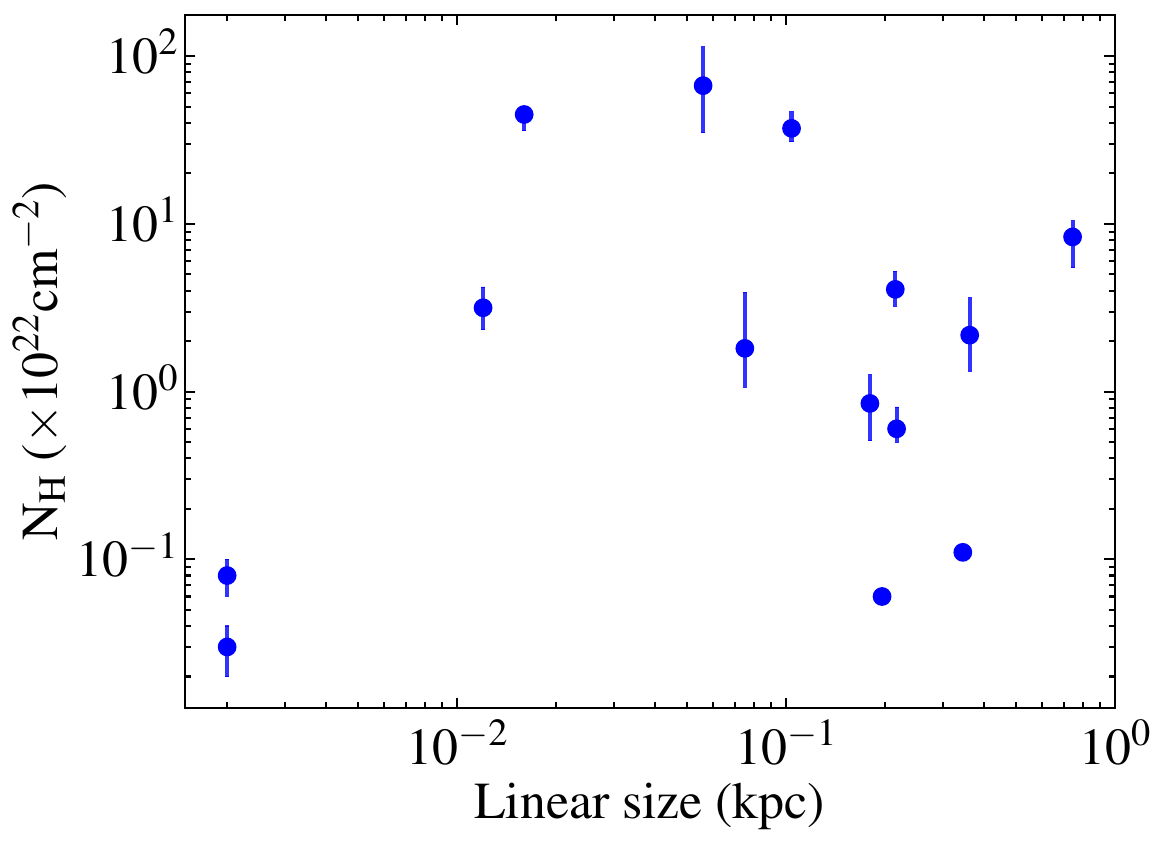}
\caption[]{The plot of linear size versus column density. The N$_{H,z}$ measurements are from {\it XMM-Netwon} and {\it Chandra} observations. The filled circles without error bars represent the upper limits of column density. The X and Y-axes are on a logarithmic scale.}
\label{figure-8}
\end{figure}

\subsection{Nature of CSOs}
Possible scenarios for the small sizes of compact steep-spectrum and peaked-spectrum sources have been summarised in Section 1 \citep[see,][for a review]{2021A&ARv..29....3O}.
A small number of CSOs are observed to have high X-ray column 
densities, which could in principle, inhibit their growth, thereby leading to compact
radio structures, as the high density interstellar medium prevents the source from expanding 
outward \citep{1984AJ.....89....5V, 1991ApJ...380...66O, Wilkinson1994, 2019ApJ...884..166S}. From simple spectral model fitting to the sources studied, we identified  
two subgroups of CSOs, namely (a) obscured  group with
N$_{H,z}$ $>$ 10$^{22}$ cm$^{-2}$ and (b) unobscured group with
N$_{H,z}$ $<$ 10$^{22}$ cm$^{-2}$. This dividing line adopted here 
is based on the value of N$_{H,z}$ $=$ 10$^{22}$ cm$^{-2}$, which divide 
AGN into X-ray obscured and X-ray unobscured \citep{2012MNRAS.426.1750M}. 
We obtained nine obscured CSOs,  
namely  J0111+3906, J1148+5924, J1347+1217, J1407+2827, J1511+0518, J1609+2641, J1945+7055, J2022+6136, and J2327+0846 \citep[see also,][]{vink2006,Bronzini2024,Siemiginowska2008,sobo2019,Sobolewska_2023,diaz2023}.

We show the intrinsic column density of the CSOs measured from {\it XMM-Newton} and {\it Chandra} observations vs their linear sizes in Figure \ref{figure-8}. For common sources, we have taken the value which has the smaller error. Two sources, J0131+5545 and J1511+0518, with errors larger than their estimated values of column density have not been considered. 
There is an indication 
of a negative correlation with high column density sources tending
to have smaller linear sizes, although the two smallest sources are in regions of low column density, which we discuss below. Excluding these two sources, half of the six sources with linear sizes smaller than 150 pc, have a column density $>37 \times 10^{22}$ cm$^{-2}$. In comparison, the maximum of column density for those larger than 150 pc is  $8.36 \times 10^{22}$ cm$^{-2}$ and the median value is  $0.85 \times 10^{22}$ cm$^{-2}$.
A similar inverse correlation  
between H{\sc i} column density and
projected linear size has been reported earlier  \citep{Pihlstrom2003A&A...404..871P,Gupta2006MNRAS.373..972G,Curran2013,Ostorero_2017} from $\lambda$21-cm absorption line studies. Although one may argue that a dense medium confines some CSOs to small dimensions, a decrease in the density of the absorbing gas with distance from the central source could also give rise to such a relationship between absorption column density and linear size. The possibility that high X-ray column densities may affect the sizes of radio sources has been suggested earlier \citep[e.g.][]{1984AJ.....89....5V, 1991ApJ...380...66O, Wilkinson1994, sobo2019,2019ApJ...884..166S,Sobolewska_2023}. 

Two sources, J1220+2916 and J1723$-$6500, with the smallest radio sizes of approximately 0.002 kpc, have N$_{H,z}$ $\leq$ 10$^{21}$ cm$^{-2}$. J1220+2916 is associated with a very nearby galaxy NGC 4278 at a redshift of 0.002. Its radio structure is dominated by a central component with weak outer components and has a flat low-frequency radio spectrum $\alpha_{408 {\rm MHz}}^{1400 {\rm MHz}} \sim 0.28$ (S$\propto \nu^{-\alpha}$). The detection of a flat radio spectrum and a bright compact core suggests that this CSO may be inclined at a smaller angle to the line of sight than most members of its class. J1723$-$6500 is also identified with a nearby galaxy NGC 6328 at a redshift of 0.014. This source was identified as the nearest GPS source by \cite{1997AJ....113.2025T}, who showed it to have two components with one of them having a flat high-frequency radio spectrum, $\alpha_{4.8 {\rm GHz}}^{8.4 {\rm GHZ}} \sim 0.3$. It has a rising low-frequency radio spectrum with $\alpha_{408 {\rm MHz}}^{1400 {\rm MHz}} \sim -0.30$. This is also a known $\gamma$-ray source \citep{2016ApJ...821L..31M}. Such sources could be intrinsically small in a low-density environment and possibly also inclined at smaller angles to the line of sight. Moreover, the inverse correlation of H{\sc i} $\lambda$21-cm absorption line column density with linear size also shows examples of small sources with low column density, possibly due to an inhomogeneous environment \citep[e.g.][]{2006MNRAS.370..738G}.

Radio observations of compact steep spectrum sources, which also include CSOs, show that sources that are associated with quasars are inclined at smaller angles to the line of sight compared with those that are associated with galaxies 
\citep[e.g.][]{Saikia1995MNRAS.276.1215S,Saikia2003PASA...20...50S}. Of the 17 sources in our sample, only one is associated with a quasar. An investigation of how orientation affects the X-ray properties of CSOs cannot be done due to the small number of sources, especially quasars. X-ray observations of more 
general CSS population are indeed needed to understand the effects of orientation on the diverse X-ray properties.

\section{Concluding remarks}
In this work, we have examined the X-ray timing and spectral characteristics of a 
sample of bona fide CSOs. For this, we have used the data from {\it Chandra}, 
{\it XMM-Newton} and {\it NuSTAR} observations that are public. We summarize the results below.

\begin{enumerate}
\item Of the 13 CSOs with {\it XMM-Newton} observations, two, J0713+4349 and J1326+3154, show
evidence of X-ray flux variations in the soft, hard and total bands with the maximum amplitude of X-ray variations being 1.17$\pm$0.27. In the case of J1326+3154, the amplitude of variations is larger in the hard band relative to the soft band. The source J1939$-$6342 was found to be a probable variable in the total band. From a model independent hardness ratio analysis, we found that both sources J0713+4349 and J1326+3154 show a HWB trend.

\item Of the 8 CSOs with {\it Chandra} observations, J0131+5545 shows
X-ray flux variability in the total energy band, although classified as a probable variable in the soft and hard bands. Among the remaining sources, J1347+1217 was found to be a probable variable in the hard and total bands, while J1723$-$6500 was found to be a probable variable in the total band. No evidence of variability was found in the 6 CSOs with {\it NuSTAR} observations. 

\item  The percentage of CSOs found to be variable at 3$\times F_{var}$ in at least one of the bands is $\sim$15, 12.5 and 0 per cent for {\it XMM-Newton}, {\it Chandra} and {\it NuSTAR} observations, respectively. Although this may increase with more sensitive and longer duration observations, the majority of CSOs do not appear to be significantly variable on a timescale of a few hours. 
It is also possible that the observations were taken during quiescent activity states and that could also contribute to the low percentage of detection of variability.


\item We found three CSOs J0111+3906, J1407+2827 and J2022+6136 to be highly obscured with a hydrogen column density $>10^{23}$ cm$^{-2}$, consistent with earlier results. 

\item Thermal components have been newly identified for the following two sources. J0111+3906 and J1945+7055 required the inclusion of a possible single thermal component
 in addition to the absorbed power-law model to better represent their 
observed X-ray spectra. 

\item For sources that have a measured black hole mass, we found no
correlation between $\Gamma$ and the Eddington ratio, $\lambda_{Edd}$. This is unlike in the case of radio-quiet AGN where a positive correlation between these two parameters possibly due to interactions of photons from the disk with the corona have been reported. The absence of a correlation for the CSOs is possibly due to contamination of the X-ray emission by jet emission in radio loud AGN.  Also, we
found a close correlation between the radio luminosity at 5 GHz and the
X-ray luminosity in the 2$-$10 keV band. This correlation suggests that 
the observed X-ray emission in them is dominated by the emission from
their relativistic jet or jet-disk coupling due to the
emission of a radiatively inefficient accretion flow, such as an ADAF suggested by the fundamental plane of blackhole activity \citep{merloni2003}.

\item We confirmed the detection of Fe K$\alpha$ line emission in J0713+4349, J1347+1217, J1407+2827, J1511+0518 and J2022+6136. The 
observed X-ray emission has a significant contribution from the disk/corona in these sources, making the scenario of radiatively inefficient accretion flow plausible.

\item We found an indication of a negative correlation between the linear size of the sources and the absorbing column density, with sources with larger N$_{H,z}$ having smaller linear sizes. This may suggest the confinement of the CSOs to small dimensions by a dense external medium. However two of the smallest CSOs which are associated with nearby galaxies are in regions of low hydrogen column density, suggesting that young CSOs may also be seen in low-density environments.  

\end{enumerate}

\section{Acknowledgements}
We thank the journal referee for several constructive criticisms and suggestions that helped us improve the manuscript. This research has made use of data obtained from the {\it Chandra} Data Archive and software provided by the {\it Chandra} X-ray Center (CXC) in
the application packages CIAO and Sherpa. This research is also based on observations obtained with {\it XMM-Newton}, an ESA science
mission with instruments and contributions directly funded by ESA member states and NASA.

\section{Data Availability}
The data from {\it Chandra}, {\it NuSTAR}, and {\it XMM-Newton} used in 
this work are from HEASARC and are in the public
domain.




\bibliographystyle{mnras}
\bibliography{reference} 



\appendix
\section{Spectral analysis result}
Tables \ref{table-5} and \ref{table-4} list the spectral parameters for a bonafide sample of CSOs from {\it XMM-Newton} and {\it Chandra} observations, respectively. \\
\begin{table*}
\centering
\caption{Results of model fits to the spectra of sources with {\it XMM-Newton}
observations. Columns show (1) J2000 name, (2) Best-fit model is highlighted with $^{*}$. The `R $^{*}$' symbol indicates the best-fitted reflection model including the Fe K$\alpha$ line. (3) Galactic hydrogen column density ($N_{H,Gal}$) in units of 10$^{22}$ cm$^{-2}$, (4) intrinsic hydrogen column density ($N_{H,ztbabs}$) in units of 10$^{22}$ cm$^{-2}$, (5) photon index (6) Normalization of power-law model (Norm$_{powerlaw}$) in units of 10$^{-4}$ photons keV$^{-1}$ cm$^{-2}$ sec$^{-1}$ at 1 keV (7),(8) Temperature in keV from Apec/blackbody model (9) Peak of Fe line emission line in keV (10) equivalent width of Fe line (11) $\chi^2$/degrees of freedom (DOF) values in the spectral fit. `f' indicates that the values are frozen to the values given in  Table (columns 5 and 7) while fitting the spectra. }
\label{table-5}
\begin{tabular}{lllllllllll} 
\hline
Name & Model &N$_{H,Gal}$ (f)  & N$_{H,ztbabs}$ & $\Gamma$ &  Norm$_{powerlaw}$ & Apec (keV) & bbody (keV) & Fe line (keV) & width (keV) &  $\chi^2$/DOF \\
(1) & (2) & (3) & (4) & (5) & (6) & (7) & (8) & (9) & (10) & (11) \\
\hline
J0029+3456 & A$^{*}$ & 0.054 & 0.85$^{+0.41}_{-0.34}$ & 1.61$^{+0.24}_{-0.27}$ & 0.47$^{+0.12}_{-0.17}$ &-&-&-&- & 1.80 \\
J0111+3906 & A & 0.055 &                        & 4.40                   & 0.009                  &-&-&-& & 2.81 \\
           & B $^{*}$ &    & 66.73$^{+47.49}_{-31.76}$ & 1.80 (f) & 0.63$^{+0.55}_{-0.33}$ &-&0.11$^{+0.03}_{-0.02}$&- &-& 0.45 \\
J0713+4349 & A &0.078 & 0.28$^{+0.09}_{-0.08}$ & 1.49$^{+0.13}_{-0.11}$ & 0.69$^{+0.09}_{-0.08}$ &-&-&-&-& 0.80 \\
           & B$^{*}$&     & 0.31$^{+0.15}_{-0.13}$ & 1.52$^{+0.14}_{-0.13}$ & 0.71$^{+0.09}_{-0.12}$ &-&-&-&& 0.76 \\
J1220+2916& A& 0.021 & 0.01$^{+0.002}_{-0.002}$ & 2.02$^{+0.02}_{-0.02}$ & 7.91$^{+0.01}_{-0.01}$ &- &-&-&-&1.21\\
          & B$^{*}$&         & 0.03$^{+0.01}_{-0.01}$ & 1.98$^{+0.03}_{-0.03}$ & 8.27$^{+0.20}_{-0.20}$ &0.77$^{+0.15}_{-0.15}$ & - &-&&1.16\\
J1326+3154 & A$^{*}$&0.012 & $<$0.11 & 1.65$^{+0.22}_{-0.19}$ & 0.21$^{+0.04}_{-0.03}$ &- &-&-&-& 1.01  \\
J1407+2827 & A& 0.015& - & 1.21$^{+0.11}_{-0.11}$ & 0.26$^{+0.02}_{-0.02}$ & &&&& 1.66\\
           & R$^{*}$ &        & 44.91$^{+1.32}_{-9.20}$ & 1.43$^{+0.15}$ & 0.86$^{+0.05}_{-0.01}$ & &-&$\sim$5.94 & - & 1.12 \\
J1443+4044 & A$^{*}$ &0.012 & 0.55$^{+1.29}_{-0.55}$ & 1.71$^{+0.33}_{-0.26}$ & 0.19$^{+0.05}_{-0.03}$ &-&-&-&-& 0.99 \\
J1511+0518&A &0.035 & <0.08 & 1.19$^{+0.26}_{-0.32}$ & 0.26$^{+0.05}_{-0.05}$ &-&-&-& & 0.77 \\
          & R$^{*}$  &       & 8.11$^{+16.19}_{-6.53}$ & 1.62$^{+0.23}$ & 0.69$^{+0.33}_{-0.24}$& 1.09 (f) &- &$\sim$5.88& $\sim$0.07&0.89 \\
J1723-6500 & A &0.060 & 0.08$\pm$0.02 & 1.88$^{+0.08}_{-0.08}$ & 1.11$^{+0.09}_{-0.08}$ &-&-&-&-& 1.21 \\
           & B$^{*}$ &       & 0.08$\pm$0.02 & 1.80$^{+0.09}_{-0.09}$ & 0.98$^{+0.09}_{-0.09}$ &0.76$^{+0.15}_{-0.22}$&-& - & - & 1.10 \\
J1939-6342 &A$^{*}$& 0.057 & $<$0.06 & 1.68$^{+0.10}_{-0.18}$ & 0.33$^{+0.06}_{-0.06}$ &-&-&-&-& 1.47  \\
J1945+7055 &A& 0.082 &1.55$^{+0.60}_{-0.50}$ & 1.13$^{+0.27}_{-0.25}$ & 0.45$^{+0.24}_{-0.15}$ &- &-&-&-& 1.35 \\
           &B$^{*}$&        &  1.65$^{+0.59}_{-0.48}$ & 1.17$^{+0.27}_{-0.24}$ & 0.47$^{+0.25}_{-0.15}$ & - & 0.18$^{+0.01}_{-0.01}$ &-&-& 0.60 \\  
J2022+6136 & A & 0.141& - & 0.49$^{+0.01}_{-0.01}$ & 0.11$^{+0.02}_{-0.02}$ &-&-&-&-& 1.83 \\
            & R$^{*}$ &       & 37.11$^{+9.66}_{-6.21}$ & 1.51$^{+0.09}_{-0.07}$ & 0.75$^{+0.21}_{-0.21}$ & - & - & $\sim$6.21 & $\sim$ 0.36 & 1.17 \\
J2327+0846 &A &0.042 & - & 2.54 & 0.86 &-&-&-&-& 5.21\\
     & R$^{*}$ &       & 8.36$^{+2.06}_{-2.89}$&   1.61$^{+0.04}_{-0.03}$ & 1.12$^{+2.06}_{-2.89}$ & 0.85$^{+0.07}_{-0.06}$ &- & $\sim$6.20 & $\sim$ 0.77 & 1.47\\
\hline
\end{tabular}
\\
\end{table*}

\begin{table*}
\centering
\caption{Results of spectral fits to the sources with {\it Chandra}
observations. Columns show (1) J2000 name, (2) Best-fit model is highlighted with $^{*}$. The `R $^{*}$' symbol indicates the best-fitted reflection model including the Fe K$\alpha$ line. (3) Galactic hydrogen column density ($N_{H,Gal}$) in units of 10$^{22}$ cm$^{-2}$, (4) intrinsic hydrogen column density ($N_{H,ztbabs}$) in units of 10$^{22}$ cm$^{-2}$, (5) photon index (6) Normalization of power-law model (Norm$_{powerlaw}$) in units of 10$^{-4}$ photons keV$^{-1}$ cm$^{-2}$ sec$^{-1}$ at 1 keV (7),(8) Temperature in keV from Apec/blackbody model (9) Peak of Fe line emission line in keV (10) equivalent width of Fe line (11) $\chi^2$/degrees of freedom (DOF) values in the spectral fit. `f' indicates that the values are frozen to the values given in Table (column 5) while fitting the spectra.}
\label{table-4}
	\begin{tabular}{lllllllllll} 
	\hline
Name & Model& N$_{H,Gal}$ (f)  & N$_{H,ztbabs}$ & $\Gamma$ &  Norm$_{powerlaw}$ & Apec (keV) & bbody (keV) & Fe line (keV) & width (keV) &  $\chi^2$/dof \\
(1) & (2) & (3) & (4) & (5) & (6) & (7) & (8) & (9) & (10) & (11) \\
\hline
J0131+5545  & A$^{*}$ & 0.250 & 0.31$^{+0.33}_{-0.29}$ & 2.37$^{+0.40}_{-0.44}$ & 4.31$^{+2.8}_{-1.62}$ &-&-&-&-& 0.78 \\ 
J0713+4349& A & 0.078    & 0.59$^{+0.20}_{-0.19}$ & 1.58$^{+0.13}_{-0.13}$ & 0.92$^{+0.02}_{-0.01}$ &-&-&-&& 1.15 \\
          & B$^{*}$ & & 0.60$^{+0.21}_{-0.10}$ & 1.61$^{+0.13}_{-0.11}$ & 0.69$^{+0.09}_{-0.08}$ &-&-&6.60$^{+0.06}_{-0.13}$&<0.221& 1.06 \\
 J1148+5924  & A & 0.019 & -                      & 0.18                   & 0.05                   &- &- & -& -& 2.34 \\
            & B$^{*}$ &       & 3.16$^{+1.05}_{-0.82}$ & 1.75$^{+0.35}_{-0.35}$ & 0.58$^{+0.46}_{-0.24}$ & 0.95$^{+0.11}_{-0.18}$& &-& & 1.18\\
J1220+2916  & A & 0.021 & - & 2.24$^{+0.06}_{-0.06}$  & 0.86$^{+0.03}_{-0.03}$ &-&-&-&-&  1.11 \\
& B$^{*}$ & & $<$ 0.02 & 2.03$^{+0.11}_{-0.07}$ & 0.76$^{+0.05}_{-0.07}$ & 0.69$^{+0.13}_{-0.11}$ &-&-&-&  0.93\\
J1347+1217  & A & 0.020 & 2.52$^{+0.53}_{-0.49}$ & 1.18$^{+0.23}_{-0.22}$ & 1.53$^{+0.65}_{-0.45}$&-&-&-&-&1.61\\
            & B$^{*}$ &       & 4.07$^{+1.16}_{-0.85}$ & 1.63$^{+0.31}_{-0.28}$ & 3.16$^{+1.95}_{-1.13}$ & - & 0.29$^{+0.08}_{-0.18}$ & 6.30$^{+0.18}_{-0.18}$ & <0.31 & 0.80 \\
J1407+2827  & A & 0.015 & - & 1.09$^{+0.11}_{-0.11}$ & 0.23$^{+0.02}_{-0.02}$ & &-&-&-&1.63 \\
            & R$^{*}$ &       & 44.09$^{+8.16}_{-12.46}$  & 1.62$^{+0.18}_{-0.08}$ & 0.97$^{+0.80}_{-0.40}$ &-&-&$\sim$ 5.92 & - & 1.38 \\
J1609+2641  & A & 0.036 & $<$0.41 & 1.32$^{+0.28}_{-0.28}$ & 0.07$^{+0.01}_{-0.01}$ &-&-& &-&1.24\\
            & B$^{*}$ &       & 2.17$^{+1.46}_{-0.85}$ & 2.27 (f) & 0.22$^{+0.06}_{-0.05}$ & - & 0.10$^{+0.08}_{-0.05}$ &-&-& 0.88 \\
J1723-6500  & A$^{*}$ & 0.060 & 0.62$^{+0.41}_{-0.33}$ & 1.89$^{+0.17}_{-0.16}$ & 1.46$^{+0.58}_{-0.41}$ & - &-&-&-& 1.64 \\ \hline
\end{tabular}
\end{table*}

\newpage
\section{Notes on individual sources}
Below, we briefly discuss the results obtained obtained both in the literature and in this work for each of the CSOs. \\

\noindent {\bf J0029+3456:} 
The source was found to show flux variations on the longer timescales ($\sim$years) based on observations taken by Einstein, ROSAT, and {\it XMM-Newton}. From absorbed power-law fits to the {\it XMM-Newton} data, \citet{2006A&A...446...87G} obtained $\Gamma$ = 1.43$^{+0.20}_{-0.19}$ and N$_{H,z}$ = $10^{+5}_{-4}$ $\times$ 10$^{21}$ cm$^{-2}$. We analysed the same dataset and obtained values of $\Gamma$ = 1.61$^{+0.24}_{-0.27}$ and 
N$_{H,z}$ = 8.5$^{+4.1}_{-3.4}$ $\times$ 10$^{21}$ cm$^{-2}$, thus in agreement with \citet{2006A&A...446...87G}. 
No significant flux variability was observed in the {\it XMM-Newton} data, similar to that reported by \citet{2006A&A...446...87G}.

\noindent{\bf J0131+5545:} \cite{2020ApJ...899..141L} analysed
the {\it Chandra} observations of this $\gamma$-ray emitting CSO taken during March - April 2019.
From absorbed power-law model fit to the {\it Chandra} observations, they obtained $\Gamma$ = 2.38$\pm$0.10 and an intrinsic absorber with an equivalent hydrogen column density of N$_{H,z} = 6.6 \times 10^{21}$ cm$^{-2}$.
Adding a thermal component ({\sc apec}) to the spectral model, the authors
obtained an upper limit to the temperature of kT = 0.09 keV.
We re-analysed the same {\it Chandra} data
with the absorbed power-law model and found a $\Gamma$ of 2.37$^{+0.40}_{-0.44}$ and N$_{H,z}$ = 3.17$^{+3.35}_{-2.92}$
$\times$ 10$^{21}$ cm$^{-2}$, similar to that
found by \cite{2020ApJ...899..141L}. From timing analysis, we found
the source to show X-ray flux variations in total band and a probable variable in the soft and hard bands. This is the first report of X-ray flux variations in the source. \\

\noindent{\bf J0713+4349:} 
This source has both {\it Chandra} and {\it XMM-Newton} observations 
\citet{Siemiginowska2016} analysed its {\it Chandra} data and found $\Gamma$ to range between 1.39 and 1.75 and N$_{H,z}$ to 
range between 0.58 to 1.02 $\times$ 10$^{22}$ cm$^{-2}$. On the other hand, 
\citet{2006MNRAS.367..928V}, from analysis of the {\it XMM-Newton} data, reported $\Gamma$ = 1.59$\pm$0.06 and N$_{H,z}$ = 0.44$\pm$0.08 $\times$ 10$^{22}$ cm$^{-2}$. 
The values of $\Gamma$ and N$_{H,z}$ obtained independently by us from simple absorbed power-law model fitting to 
both the {\it Chandra} and {\it XMM-Newton} spectra are similar to that
of \cite{Siemiginowska2016} and \cite{2006MNRAS.367..928V}. We also found an ionised Fe emission line at 6.60$^{+0.06}_{-0.13}$ keV with equivalent width $<$0.221 keV from {\it Chandra} data, which is in agreement with \cite{Siemiginowska2016}. We found the source to show flux 
variations in all three bands in the {\it XMM-Newton} observations, and also a HWB trend. However, the source did not show flux variability in the {\it Chandra} observation. 

\noindent{\bf J1148+5924:}
This CSO is reported to be a $\gamma$-ray emitter by \citet{2020A&A...635A.185P}. From an analysis of the {\it Chandra} observations in 2009, \cite{2021ApJ...922...84B} found the spectrum to be well fit by a combination of ionized thermal plasma with a temperature of 0.8 keV and an absorbed power-law model with $\Gamma$ = 1.4$\pm$0.4, and $N_{\rm H,z} = 2.4\pm0.7 \times 10^{22}$ cm$^{-2}$. They also detected the Fe K$\alpha$ line at 6.5$\pm$0.1 keV with an equivalent width of 1.0$^{+0.9}_{-0.5}$ keV with the spectrum being binned to have 5 counts per energy bin. Recently, \cite{Bronzini2024} revealed the presence of a multi temperature thermal component dominating the soft X-ray spectrum using archival {\it Chandra} and {\it NuSTAR} observations with $\Gamma$ = 1.92$^{+0.34}_{-0.33}$ and a moderate $N_{\rm H,z} = 3.49^{+1.28}_{-1.04} \times 10^{22}$ cm$^{-2}$.
In this work, we re-analysed the {\it Chandra} and {\it NuSTAR} data of the source. The application of the absorbed power-law and an ionised thermal plasma component (Equation 6), we obtained $\Gamma$ = 
1.75$^{+0.35}_{-0.35}$, 
$N_{\rm H,z} = 3.16^{+1.05}_{-0.82} \times 10^{22}$ cm$^{-2}$, and a kT of 
0.95$^{+0.11}_{-0.18}$ keV. The Fe K$\alpha$ emission line was also detected. Our spectral analysis results are in agreement with that published by \citet{she_2017} and \citet{Bronzini2024}. 
The source is non-variable in the X-ray band.

\noindent{\bf J1220+2916:}
This source exhibits a LINER and Seyfert nucleus \citep{Younes2010}. It has been observed with both {\it Chandra} and {\it XMM-Newton}. \citet{2014MNRAS.443...72J} obtained values of $\Gamma$ = 2.06$\pm$0.01 and $N_{\rm H,z} = 0.02\pm0.01 \times 10^{22}$ cm$^{-2}$ from the {\it XMM-Newton} observations.  
Using the {\it XMM-Newton} data, we obtained a good fit with $\chi^2$ 1.21 with model-A and 1.17 for model-B. Using thermal diffuse emission model, we obtained $\Gamma$ = 1.98$^{+0.03}_{-0.03}$, kT = 0.77$^{+0.15}_{-0.15}$ keV and $N_{\rm H,z} = 0.01^{+0.003}_{-0.003} \times 10^{22}$ cm$^{-2}$. We obtained similar parameters from {\it Chandra} data analysis. Our independent analysis using both an absorbed power-law (model-A) and an absorbed power-law
plus a thermal component (model-B) from {\it XMM-Newton} observations are in agreement with the results available in the literature \citep{2014MNRAS.443...72J,2014A&A...569A..26H}. There is no statistical difference between model-A and model-B with a $p$-value of 0.26. However, we have evidence of a thermal plasma component of 0.77 keV using {\it XMM-Newton} data, consistent with findings from {\it Chandra} observation by \cite{Younes2010}. They obtained similar results with kT $\sim$0.6 keV and $N_{\rm H,z}$ of 0.01 $\times$ 10$^{22}$ cm$^{-2}$.
No significant flux variability was observed in the {\it XMM-Newton} and {\it Chandra} datasets.


\noindent{\bf J1326+3154:}
This source was observed by {\it XMM-Newton} on 05 December 2007. \citet{2009A&A...501...89T} analysed this data and reported $\Gamma$ = 1.7$\pm$0.2 and $N_{\rm H,z} = 1.2^{+0.6}_{-0.5} \times 10^{21}$ cm$^{-2}$. From spectral analysis of the same dataset, we found a $\Gamma$ of 1.65$^{+0.22}_{-0.19}$ and an upper limit of $N_{\rm H,z} = 1.1 \times 10^{21}$ \rm {cm}$^{-2}$ which is similar to that found by \cite{2009A&A...501...89T}. Furthermore, we found
the source to show significant flux variations in all the bands, with
the variations in the hard band larger than the soft band. This is
the first detection of X-ray flux variability in this source. Also, the
source was found to exhibit a HWB trend. \\

\noindent{\bf J1347+1217:}
Extended X-ray emission of the order of $\sim$20 kpc was noticed in the {\it Chandra} data of this CSO and was explained due to thermal emission from the galaxy halo \citep{2008ApJ...684..811S}. 
From the absorbed power-law model fit to the spectrum, the authors found a $\Gamma$ = 1.10$^{+0.29}_{-0.28}$ and a large absorbing column density with $N_{\rm H,z} = 2.54^{+0.63}_{-0.58} \times$ $10^{22}$ \rm{cm}$^{-2}$. These are nearly similar to the values obtained by \cite{2018ApJ...868...10L}, such as 
$\Gamma$ = 1.69$^{+0.30}_{-0.20}$, $N_{\rm H,z} = 3.23\pm0.52 \times 10^{22}$ cm$^{-2}$.
\citet{2013ApJ...777...27J}
reported the detection of a Fe K$\alpha$ line at 6.42$^{+0.07}_{-0.08}$ keV.
From simple absorbed power-law fit, we found $\Gamma$ = 1.18$^{+0.23}_{-0.22}$ and an intrinsic neutral hydrogen
column density of $N_{\rm H,z} = 2.52^{+0.53}_{-0.49} \times 10^{22}$ \rm{cm}$^{-2}$ similar to that reported by \cite{2008ApJ...684..811S}.
Additionally, to account for the thermal emission in the soft band, we used
a simple blackbody model, and arrived at a temperature 
of kT = 0.29$^{+0.08}_{-0.18}$ keV
and detected the Fe K$\alpha$ line at 6.30$\pm$0.18 keV, with an
equivalent width $<$0.3 keV. \cite{2013ApJ...777...27J} reported an equivalent width of the identified Fe K$\alpha$ line to be 0.20$^{+0.15}_{-0.12}$ keV, which is in agreement with the upper limit obtained in this work. Our spectral analysis results with $\Gamma$=1.63$^{+0.31}_{-0.28}$ and $N_{\rm H,z} = 4.07^{+1.16}_{-0.85}
\times 10^{22}$ cm$^{-2}$ are also similar to that found by \cite{2014ApJ...787...61L}. From the timing analysis, we found the source to be probable variable in the hard and total energy bands. This is the first report of the X-ray flux variations in the source.

\noindent{\bf J1407+2827:}
From the ASCA and {\it XMM-Newtwon} observations, this source was found to have a flat X-ray spectrum with a strong Fe K$\alpha$ line with equivalent width of about 0.9 keV \citep{2004A&A...421..461G}. The analysis of the joint {\it NuSTAR}, {\it Chandra}, and {\it XMM-Newton} spectrum of this object was carried out by \citet{2019ApJ...884..166S} who reported
a high obscuration with $N_{\rm H,z} \approx 3\times 10^{23}$ \rm{cm}$^{-2}$ and the primary X-ray emission characterised by $\Gamma$ = 1.45$\pm$0.11. 
From an absorbed power-law fitting of 
{\it Chandra}, {\it XMM-Newton} data, we found the source to have a 
flat spectral index of $\Gamma$ = 1.16 $\pm$ 0.12 and
1.21 $\pm$ 0.11 respectively. Also, Fe K$\alpha$ line is detected in both 
the {\it XMM-Newton} and {\it Chandra} observations. The width of Fe K$\alpha$ line was constrained to be $\sim$0.9 keV using {\it XMM-Newton} data, which is in agreement with that found by \cite{2004A&A...421..461G}. We also included {\it NuSTAR} spectrum for the fitting and applied {\sc borus} model as distant reflection model. We found the N$_{H,z}$ =  4.49$^{+0.13}_{-0.92}$ $\times$ 10$^{23}$ cm$^{-2}$, $\Gamma$ = 1.43$^{+0.15}_{-0.13}$ and covering factor = 0.79$^{+0.21}_{-0.23}$, consistent with the results of \cite{2019ApJ...884..166S}. 
The peak and EW of the Fe K$\alpha$ emission line with respect to the total continuum are found to be 5.94 keV and $\sim$0.69 keV.  
The source exhibited no significant flux variability in the {\it XMM-Newton} and {\it Chandra} datasets. 

\noindent{\bf J1443+4044:}
This source was 
observed by {\it XMM-Newton} on 18 January 2019 for a duration of 33 ksec. It was studied by \citet{Chartas_2021} who reported the detection of ultrafast outflows with two velocity components. By fitting an absorbed power-law model, we obtained 
a $\Gamma$ of 1.71$^{+0.33}_{-0.26}$ which is in agreement with 1.80$\pm$0.12 reported by \citet{Chartas_2021}. The estimated intrinsic neutral hydrogen column density is N$_{H,z}$ =  0.55$^{+1.29}_{-0.55}$ $\times$ 10$^{22}$ cm$^{-2}$. The source showed no significant flux variability in the {\it XMM-Newton} observation. From the timing analysis of the same observation, no flux variability could be 
detected. 

\noindent{\bf J1511+0518:}
This source was observed by {\it XMM-Newton} on 15 August 2018 for a duration of 9.6 ksec. This source was also observed by {\it Chandra} for about 2 ksec.
Fitting the data with the sum of an 
absorbed power-law and reflection model ({\sc pexrav}),
\cite{Siemiginowska2016} estimated values of $\Gamma$ = 3.8$^{+0.3}_{-0.1}$ and $N_{\rm H,z} = 3.8^{+4.0}_{-1.3} \times 10^{23}$ cm$^{-2}$. 
Its {\it NuSTAR} observations were recently studied by \cite{Sobolewska_2023} who constrained the photon index to be $\sim$ 1.6-1.7,  kT $\sim$ 1 keV and a neutral hydrogen column density of $\sim$ 10$^{23}$ cm$^{-2}$ along the torus line of sight. The study suggested the presence of thermal components and dusty torus around the active nucleus. 
We re-analysed the {\it XMM-Newton} data in this 
work and found the source not to show any flux variations. From an 
absorbed power-law model fit to the {\it XMM-Newton} data, we obtained a 
$\Gamma$ of 1.19$^{+0.26}_{-0.32}$ with the column density < 1 $\times 10^{22}$ cm$^{-2}$. Addition of a thermal component to the absorbed power-law model, we found $N_{\rm H,z} = 3.74^{+5.57}_{-2.02} \times 10^{22}$ cm$^{-2}$ 
with a kT of 0.25$^{+0.09}_{-0.08}$ keV. Adding {\sc Pexrav} model to it, we obtained a photon index $<$3.2 similar to that of 
\cite{Siemiginowska2016}. We carried out the joint fitting of {\it XMM-Newton} and {\it NuSTAR} data with {\sc Borus model}. We found the better $\chi^2$ value of 0.89 and hence adopted it as the final model. We found $\Gamma$ = 1.62$^{+0.23}$, $N_{\rm H,z} = 8.11^{+16.19}_{-6.53} \times 10^{22}$ cm$^{-2}$ consistent with that estimated by \citet{Sobolewska_2023}. We found the peak and EW of the Fe K$\alpha$ emission line with respect to the total continuum at $\sim$ 5.88 keV and $\sim$0.07 keV. This source is found to be non-variable from the timing analysis.

\noindent{\bf J1609+2641:}
The X-ray properties of  this CSO have been studied by
\citet{2009A&A...501...89T} and \citet{Siemiginowska2016}.
\cite{2009A&A...501...89T} reported the detection of Fe K$\alpha$ line in
this source and $\Gamma$ = 0.4$\pm$0.3. However, 
\cite{Siemiginowska2016}
found the source to have a companion separated by about 13 arcsec and
the results reported by \cite{2009A&A...501...89T} were possibly
affected by the contamination of the companion object. \cite{Siemiginowska2016} 
obtained a value of $\Gamma$ = 1.4$\pm$0.1 and they did not find any evidence of
the Fe K$\alpha$ line.
From the absorbed power-law model fit to the {\it Chandra} observation,
we obtained a $\Gamma$ of 1.32$\pm$ 0.28. We found an upper limit
to the neutral hydrogen column density of 0.41 $\times$ 10$^{22}$ \rm{cm}$^{-2}$.
On addition of a thermal component to the absorbed power
law model, we found $N_{\rm H,z} = 2.17^{+1.46}_{-0.85}$ $\times$ 10$^{22}$ cm$^{-2}$. The thermal component has not been investigated in the literature due to fewer photon counts. When we froze $\Gamma$ at its best-fit value, we found a possible thermal component with kT of 0.10$^{+0.08}_{-0.05}$ keV. The X-ray light curve of the source has not revealed any significant flux variability.

\noindent{\bf J1723$-$6500:}
This source is a $\gamma$-ray emitting CSO \citep{2016ApJ...821L..31M}.
From the first X-ray observations carried out by {\it Chandra}
on 09 November 2011, it was found to have an extended X-ray emission. The X-ray spectrum can be described by an absorbed power-law
model with  $\Gamma$ = 1.6$\pm$0.2 and 
$N_{\rm H,z} = 0.08\pm0.07 \times 10^{22}$ cm$^{-2}$
\citep{Siemiginowska2016}. \citet{2018A&A...612L...4B} carried out a detailed analysis of multi-epoch X-ray data of this object and derived $\Gamma$ = 1.78$^{+0.10}_{-0.99}$, 
$N_{\rm H,z} = (3-7) \times 10^{21}$ cm$^{-2}$ which is consistent with the simultaneous fitting of {\it XMM-Newton} and {\it NuSTAR} data studied by \cite{Bronzini2024}. 
From an independent analysis of the {\it XMM-Newton} and {\it Chandra} and {\it NuSTAR} observations reported here, our values of $\Gamma$, N$_{H,z}$ and diffuse plasma temperature (see Table 4 and Table 5) are similar to the values found by previous works. From timing analysis, we found the source as a probable variable in the total band in the {\it Chandra} data.

\noindent{\bf J1939$-$6342:}
\cite{2003A&A...401..895R} studied this CSO using BeppoSAX observations and found it to be a possible Compton-thick source with $N_{\rm H,z} > 2.5 \times 10^{24}$ cm$^{-2}$. They also reported the detection of Fe K$\alpha$ line and a reflection hump.
However, {\it Chandra}, observations resulted in a 3$\sigma$ upper limit of the 
Fe K$\alpha$ line equivalent width $<$ 0.96 keV and measured a low absorption 
column density of $N_{\rm H,z} = 0.08^{+0.07}_{-0.06} \times 10^{22}$ cm$^{-2}$ \citep{Siemiginowska2016}.
From the absorbed power-law fit to the {\it XMM-Newton} spectrum, we could 
not constrain the absorbing column density. We found a value of $\Gamma$ of 1.68$^{+0.10}_{-0.18}$ and $N_{\rm H,z} < 0.06 \times 10^{22}$ cm$^{-2}$.
Our results based on {\it XMM-Newton} are in agreement with those obtained by
\cite{Siemiginowska2016} and \cite{sobo2019} based on {\it Chandra} and {\it XMM-Newton} data. 
From the analysis of the 
{\it XMM-Newton} data acquired on 01 April 2017, we considered this source as a probable variable in the total band based on the criteria adopted in this work. \\

\noindent{\bf J2022+6136:}
This source has been observed by {\it Chandra} and {\it XMM-Newton}. On 
analysis of the data from {\it Chandra} observed on 04 April 2011, 
\cite{Siemiginowska2016} found a flat X-ray spectrum with $\Gamma$ of 0.8$^{+0.3}_{-0.2}$, indicating the source to be a possible Compton thick CSO. Moreover, using
an absorbed power-law model along with an unabsorbed reflection ({\sc pexrav}), 
they obtained a soft $\Gamma$ = 3.3$\pm$ 0.3 and $N_{\rm H,z} > 9.5 \times 10^{23}$ cm$^{-2}$. We fitted an absorbed power-law model to the {\it XMM-Newton} and found $\Gamma$ = 0.49$\pm$0.01. Inclusion of a Gaussian component to the absorbed power-law model, we found Fe K$\alpha$ line at 6.16$^{+0.50}_{-0.27}$ keV with an equivalent width of 0.85$^{+0.91}_{-0.38}$ 
keV. In order to account for the reflection component, we used the {\sc borus} model with {\it XMM-Newton} and {\it NuSTAR} data. We found the photon index $\Gamma$= 1.51$^{+0.09}_{-0.07}$ and $N_{\rm H,z} = 3.71^{+0.96}_{-0.21} \times 10^{23}$ cm$^{-2}$, and covering factor = $\sim$ 0.73 which is in agreement with \cite{Sobolewska_2023}. The equivalent width for Fe K$\alpha$ line emission is found to be $\sim$ 0.36 keV. This source is found to be non-variable by the timing analysis.

\noindent{\bf J2327+0846:} 
This CSO is known as a Type 2 Seyfert and possibly a changing look AGN \citep{diaz2023}. This source has been extensively studied for its X-ray properties 
\citep{2005A&A...442..185B,2017MNRAS.467.4606G, 2020ApJ...897....2T,2020ApJ...894...71Z}.
\cite{2017MNRAS.467.4606G} studied this object using {\it NuSTAR}, {\it Suzaku} and {\it Swift} datasets and reported the CSO to be a candidate Compton thick source with $N_{\rm H,z} > 2 \times 10^{24}$ cm$^{-2}$. However, \cite{2020ApJ...897....2T} reported a column density of
$N_{\rm H,z} \sim 2 \times$ 10$^{23}$ cm$^{-2}$, when the spectra was fitted
with the XCLUMPY model. 
We combined both the observations from {\it XMM-Newton} and {\it NuSTAR} and found $\Gamma$ = 1.61$^{+0.04}_{-0.03}$ and N$_{\rm H,z}$ = 8.36$^{+2.06}_{-2.89}$ $\times$ 10$^{22}$ cm$^{-2}$. The timing analysis suggested this object to be non-variable.




\bsp	
\label{lastpage}

\end{document}